\documentclass[aps,prb,twocolumn,longbibliography,floatfix]{revtex4-2}

\usepackage[bookmarks, colorlinks=false]{hyperref}
\hypersetup{colorlinks=true, allcolors=blue}
\usepackage{placeins}

\usepackage{mathtools}

\usepackage{chngcntr}
\usepackage{booktabs}
\usepackage{tabularx}
\usepackage{xcolor}
\usepackage{dcolumn}
\usepackage{bm}
\usepackage{siunitx} 

\newcolumntype{K}[1]{>{\centering\arraybackslash}p{#1}}

\usepackage{multirow}

\definecolor{commentcol}{rgb}{0.5, 0.2, 0.0}

\usepackage[caption=false]{subfig}

\begin{document}

\title{An empirical potential to simulate helium and hydrogen in highly irradiated tungsten}

\author{Samanyu Tirumala}
\email{samanyu.tirumala@ukaea.uk}
\affiliation{UK Atomic Energy Authority, Culham Campus, OX14 3DB Abingdon, UK}
\affiliation{Department of Engineering Science, University of Oxford, Parks Road, OX1 3PJ Oxford, UK}

\author{Daniel R. Mason}
\affiliation{UK Atomic Energy Authority, Culham Campus, OX14 3DB Abingdon, UK}

\author{Oliver Shattock}
\affiliation{School of Physics, Engineering and Technology, University of York, Heslington, York YO10 5DD, UK}

\author{Duc Nguyen-Manh} 
\affiliation{UK Atomic Energy Authority, Culham Campus, OX14 3DB Abingdon, UK}
\affiliation{Department of Materials, University of Oxford, Parks Road, OX1 3PH Oxford, UK}

\author{Felix Hofmann}
\email{felix.hofmann@eng.ox.ac.uk}
\affiliation{Department of Engineering Science, University of Oxford, Parks Road, OX1 3PJ Oxford, UK}

\author{Max Boleininger}
\email{max.boleininger@ukaea.uk}
\affiliation{UK Atomic Energy Authority, Culham Campus, OX14 3DB Abingdon, UK}

\begin{abstract}
\noindent Materials used in commercial D-T fusion reactors will be exposed to irradiation and a mixture of helium and hydrogen plasma. Modeling the microstructural evolution of such materials requires the use of large-scale molecular dynamics simulations. The focus of this study is to develop a fast EAM potential for the interactions among the three elements (W, H, and He), fitted to accurately reproduce both the ab initio formation energies and relaxation volumes of small defect clusters containing light gases within tungsten. The potential enables the study of tungsten under irradiation and in the presence of light gases. To demonstrate the utility of the potential, we construct a thermodynamically motivated model for predicting the energetics of light-gas-filled voids. The model is then validated through molecular dynamics simulations with our new potential.

\end{abstract}

\date{\today}

\maketitle


\section{Introduction}

\noindent To build a commercially viable fusion reactor, it is crucial to develop a rigorous understanding of the long-term evolution of the materials that make up the reactor. Especially plasma-facing components (PFCs) will be subjected to extreme operating conditions, including irradiation by high-energy neutrons and implantation of helium and lower-energy fluxes of hydrogen isotopes \cite{ueda2017baseline}. Tungsten is a candidate PFC material \cite{luo2023research, wurster2013recent, ueda2014research, rieth2013recent, marian2017recent} due to its excellent properties, such as a high melting point, low sputtering yield \cite{saidoh1983low}, and high thermal conductivity. Still, significant microstructural changes are expected to occur during reactor operation due to accumulation of irradiation damage, such as the formation of gas bubbles containing helium and hydrogen isotopes \cite{ueda2009effects, roth2011hydrogen}. Furthermore, light gases have been shown to bind to dislocations, thereby impeding their motion and leading to both hardening and embrittlement of the material \cite{das2019hardening, li2020three}. Consequently, the presence of light gases in tungsten not only degrades its performance and longevity \cite{ueda2017baseline, luo2023research}, but also enhances tritium retention, ultimately reducing the fuel efficiency of the reactor.

Given the significant costs and challenges associated with building and testing reactors, and the relative scarcity of experimental data of materials behavior under relevant conditions, computational models are essential to assess material durability under fusion conditions. The primary effect of irradiation is the generation of defects at the atomic scale, such as interstitials and vacancies 
which coalesce into voids \cite{surh2008void} and dislocation loops \cite{sand2013high}, respectively. The properties of these nanoscale defects are typically studied with first-principles methods, such as density functional theory (DFT). Due to the high computational demand of DFT, its applicability is limited to systems comprising between hundreds and thousands of atoms \cite{mason2019rv,ma2019universality, ma2020multiscale}. Transmission electron microscopy (TEM) studies have shown that the defects induced by heavy irradiation and implantation can organize into microstructural features beyond the micrometer scale \cite{el2014situ, ipatova2021situ}. Molecular dynamics (MD) simulations can be employed to simulate such structures. To maintain DFT-level accuracy, the interatomic potentials used in the MD simulation must be fitted to replicate properties derived from DFT or relevant experimental properties. Recent advancements in machine learning potentials (MLPs) \cite{drautz2019atomic, deringer2019machine, mishin2021machine, goryaeva2021efficient, cusentino2024} have resulted in interatomic potentials with accuracy approaching DFT. However, evaluation of these potentials presently remains orders of magnitude slower than classical potentials such as the embedded atom method (EAM) \cite{LysogorskiyYury2021Piot}.

In this work, we present the development of a classical potential for the ternary W-H-He system for radiation damage simulations. We selected an EAM potential due to its balanced combination of computational efficiency and accuracy. EAM potentials are widely recognized for their suitability in modeling metallic systems \cite{ackland1987improved, marinica2013interatomic, bonny2013interatomic, lin2008computational} and have demonstrated the ability to predict properties of irradiated microstructures consistent with experimental observations \cite{mason2017,mason2021parameter, warwick2023dislocation, feichtmayer2024fast}. We developed EAM potentials tailored to reproduce the ordering of activation and binding energies of radiation defects containing light gases, enabling a qualitative description of the emergent complex behaviour. In our study, we focus on two key properties of macroscopic significance: the binding energy of light gases to vacancy-type defects and the relaxation volume of such defects. 

Experiments have demonstrated a strong spatial correlation between the irradiation dose depth-profile and the retention of helium and hydrogen \cite{schwarz2018influence, karcher2023influence}. Complementary DFT studies \cite{nguyen2015trapping, yang2018energetics, heinola2010hydrogen} indicate that this retention is primarily governed by the trapping of these gases within irradiation-induced void defects. Although binding to dislocation loops is possible, De Backer et al. demonstrated that at finite temperature the amount of hydrogen trapped at loops is relatively small, suggesting that retention must instead be dominated by other trapping mechanisms - namely vacancies and voids.

At low irradiation doses, small vacancy clusters are the dominant void-like structures. With increasing dose, temperature, and light gas concentration, experiments have shown an increase in void sizes \cite{ipatova2021situ, harrison2017study, miyamoto2015systematic}. Since vacancies and voids play a significantly role as trapping sites for light gases, it is essential for our EAM potential to offer an accurate description of the vacancy-gas and void-gas interactions.

Irradiation-induced swelling has been well documented in both experimental studies \cite{toloczko1996irradiation, matolich1974swelling, hofmann2015lattice, feichtmayer2024fast, garner1988irradiation, matthews1988irradiation} and MD simulations \cite{feichtmayer2024fast, reali2025atomistic}. Dudarev \textit{et al.} \cite{dudarev2017elastic, dudarev2018multi} and Reali \textit{et al.} \cite{reali2021macroscopic} have shown that defect relaxation volumes are primary contributors to the macroscopic swelling mechanism and are key parameters in describing the elastic field interactions between defects. The presence of helium significantly amplifies this swelling, particularly under high flux and dose conditions, where the formation of large helium bubbles can lead to surface fractures in the material \cite{allen2020key}. We therefore place emphasis on accurately modeling the relaxation volumes of clusters of light gases trapped within both the lattice and vacancies.

Various W-H-He potentials exist in the literature, including fully fitted models such as the potentials by Li \textit{et al.} \cite{li2025analytical} and Bonny \textit{et al.} \cite{bonny2014binding}, as well as hybrid approaches that merge existing potentials such as the potential by Yang \textit{et al.} \cite{yang2018effect}, which combines the W-W by Marinica \textit{et al.} \cite{marinica2013interatomic}, W-H by Wang \textit{et al.} \cite{wang2017embedded}, and W-He by Juslin \textit{et al.} \cite{juslin2013interatomic}. In the following, we shall refer to the potentials according to the first author of the corresponding publication. The Bonny potential was not fitted to reproduce the gaseous properties of hydrogen and helium, leading to inaccuracies when applied to bubbles. In contrast, both the Yang and Li potentials effectively describe hydrogen-helium interactions within the tungsten lattice and gaseous phases, providing a more reliable description of the W-H-He system. Since the Li potential was explicitly fitted to the W-H-He system, it offers a more accurate representation of defect properties compared to the Yang potential. However, these potentials were primarily fitted to capture the energetics of defects in the W-H-He system without considering the relaxation volumes. As a result, they tend to overestimate these volumes. Additionally, inaccuracies in the predictions of tungsten surface energies lead to an imprecise description of voids. This highlights the necessity of developing a new potential that addresses these limitations and provides a more comprehensive model for studying common radiation defects in the W-H-He system.

Here, we describe the development of a new W-H-He potential building upon the W-H potential developed by Mason \textit{et al.} \cite{mason2023empirical}, which accurately reproduces DFT-calculated binding energies and relaxation volumes of typical small irradiation-induced defects containing hydrogen. Furthermore, the potential provides a good description of tungsten surfaces, enabling accurate modeling of voids within the lattice. This alignment with our modeling goals makes it a suitable choice. We extend this potential to also include helium interactions, allowing for the simulation of both hydrogen and helium in irradiated tungsten. In the remainder of this paper, we compare the newly developed potential with existing models in the literature. Specifically, we select the Li potential \cite{li2025analytical} and the hybrid approach used by Yang \cite{yang2018effect} for comparison.


\section{Fitting Method}

\noindent We employ the generalized Finnis-Sinclair potential \cite{finnis1984simple}. In this formulation, the total energy  \(E_a\) of an atom $a$ depends on the interatomic distances $r_{ab}$ to its neighboring atoms \(b \in \text{nebs}\), and the respective atomic species, $\alpha$ and $\beta$: 
\begin{equation}
    E_a = \sum_{b \in \text{nebs}} \phi_{\alpha \beta}(r_{ab}) +
    F_{\alpha} \left(\sum_{b \in \text{nebs}} \rho_{\alpha \beta}(r_{ab}) \right).
    \label{eq:fs energy}
\end{equation}
The potential \(\phi_{\alpha \beta}\) represents pairwise interactions between atom \(a\) of species \(\alpha\) and atom \(b\) of species \(\beta\). The function \(\rho_{\alpha \beta}\) represents the electron density contribution from atom \(b\) of species \(\beta\) to atom \(a\) of species \(\alpha\). The embedding function \(F_{\alpha}\) represents the many-body interaction energy required to place an atom \(a\) of species \(\alpha\) into the electron density generated by the surrounding atoms. We parameterize the functions using splines and continuous expressions, with details given in Sec.~\ref{sec:potpar}.

Zhou \textit{et al.} \cite{zhou2021enabling} demonstrated that allowing the electron density to take negative values leads to a significantly more accurate representation of the DFT properties for metallic hydrogen-helium systems. Accordingly, we allow the electron density contributions \(\rho_{\alpha \beta}\) to take on negative values during fitting, which can be interpreted as species \(b\) inducing a localized reduction in electron density around species \(a\). This generalized form of the Finnis-Sinclair model is implemented in the MD simulator \textsc{Lammps} \cite{thompson2022lammps} under the \texttt{pair\_style eam/he} \cite{zhou2021enabling} option.

Since the Mason W-H potential follows the \texttt{eam/alloy} formulation, we needed to extend the embedding functions of tungsten and hydrogen to accommodate negative electron density values. In the Mason potential, an embedding function is generally defined as:
\begin{equation} 
F_\alpha(\rho) = A_\alpha \sqrt{\rho} + s_\alpha(\rho),
\end{equation}
where $s_\alpha(\rho)$ is a set of quintic spline functions constrained to be zero at $\rho = 0$. For the extension to negative electron densities, we adopt the following modification for the tungsten and hydrogen electron densities:
\begin{equation} F_\alpha(-\rho) = - F_\alpha(\rho)
\end{equation}

\subsection{Optimization Algorithm}
\noindent Let a function \(f\) to be fitted be defined by a parameter set \(\mathbf{x}\), which may include knot parameters, function coefficients, or other defining parameters. To identify the optimal parameters, we introduce a single-valued loss function, with the parameter set minimizing this function representing the best fit. Here, the loss is given by the difference in defect properties with respect to DFT data. For a given parameter set  \(\mathbf{x}\) and given sets of defect configurations \(\{D_i\}\), defect reactions \(\{R_j\}\), and gaseous configurations \(\{G_k\}\), we define the loss function as:
\begin{equation}
\begin{aligned}
    \mathcal{L}(\mathbf{x}) = \phantom{+}& 
    \sum_i \alpha_i \left |\Delta E_\mathrm{f}(\mathbf{x},D_i)\right| 
    + \beta_i \left |\Delta \Omega(\mathbf{x},D_i)\right| \\
    +&\sum_j \gamma_j \left |\Delta E_\mathrm{b}(\mathbf{x}, R_j)\right| 
    + \sum_k \delta_k \left |\Delta E(\mathbf{x},G_k) \right|
    \label{eq:loss_equation}
\end{aligned}
\end{equation}
The first term \(\left |\Delta E_\mathrm{f}(\mathbf{x},D_i)\right|\) represents the loss due to differences in the formation energy of defect \(D_i\); this term specifically only considers single-helium defects to ensure accurate description of the migration barrier, as well as the formation energies in tetrahedral and octahedral sites. The second term \(\left |\Delta \Omega(\mathbf{x},D_i)\right|\) represents the loss due to differences in the relaxation volume of defect \(D_i\). The third term \(\left |\Delta E_\mathrm{b}(\mathbf{x}, R_j)\right|\) represents the loss due to differences in binding energy for a given reaction \(R_j\), which includes scenarios such as the binding energy released when two interstitial helium atoms cluster together. The final term \(\left |\Delta E(\mathbf{x}, G_k)\right|\) represents the loss due to differences in the total energy of gaseous configurations \(G_k\), ensuring that the properties of gaseous states for light gases are accurately captured. 

The loss function aims to balance accuracy between defect properties and gaseous properties. To align with the objectives of this work, greater emphasis is placed on defect binding energies and their associated relaxation volumes. As a result, the weights \( \gamma \) and \( \delta \), corresponding to these defect properties, are generally assigned larger values than the weights \( \alpha \) and \( \beta \), which correspond to gaseous properties and secondary contributions. To avoid overfitting the potential to only defect-related quantities, \( \alpha \) and \( \beta \) are still chosen to be of the same order of magnitude as \( \gamma \) and \( \delta \), ensuring that all property types influence the fitting process.

Furthermore, certain properties are considered more important than others, which is reflected in the variation of the weights, i.e. it is not necessarily the case that \( \alpha_i = \alpha_j \). For example, greater importance is assigned to binding energies on the order of {1\,eV} or lower, as these are more likely to influence defect dynamics at temperatures relevant to molecular dynamics (MD) simulations. In contrast, higher binding energies, e.g. above {2\,eV}, correspond to processes that are thermally unlikely to reverse and are therefore weighted less heavily. The weightings were tuned through a combination of trial and error and by evaluating whether the resulting potential accurately reproduced the desired set of physical properties.

\begin{figure}[t]
    \centering
    \includegraphics[width=\columnwidth]{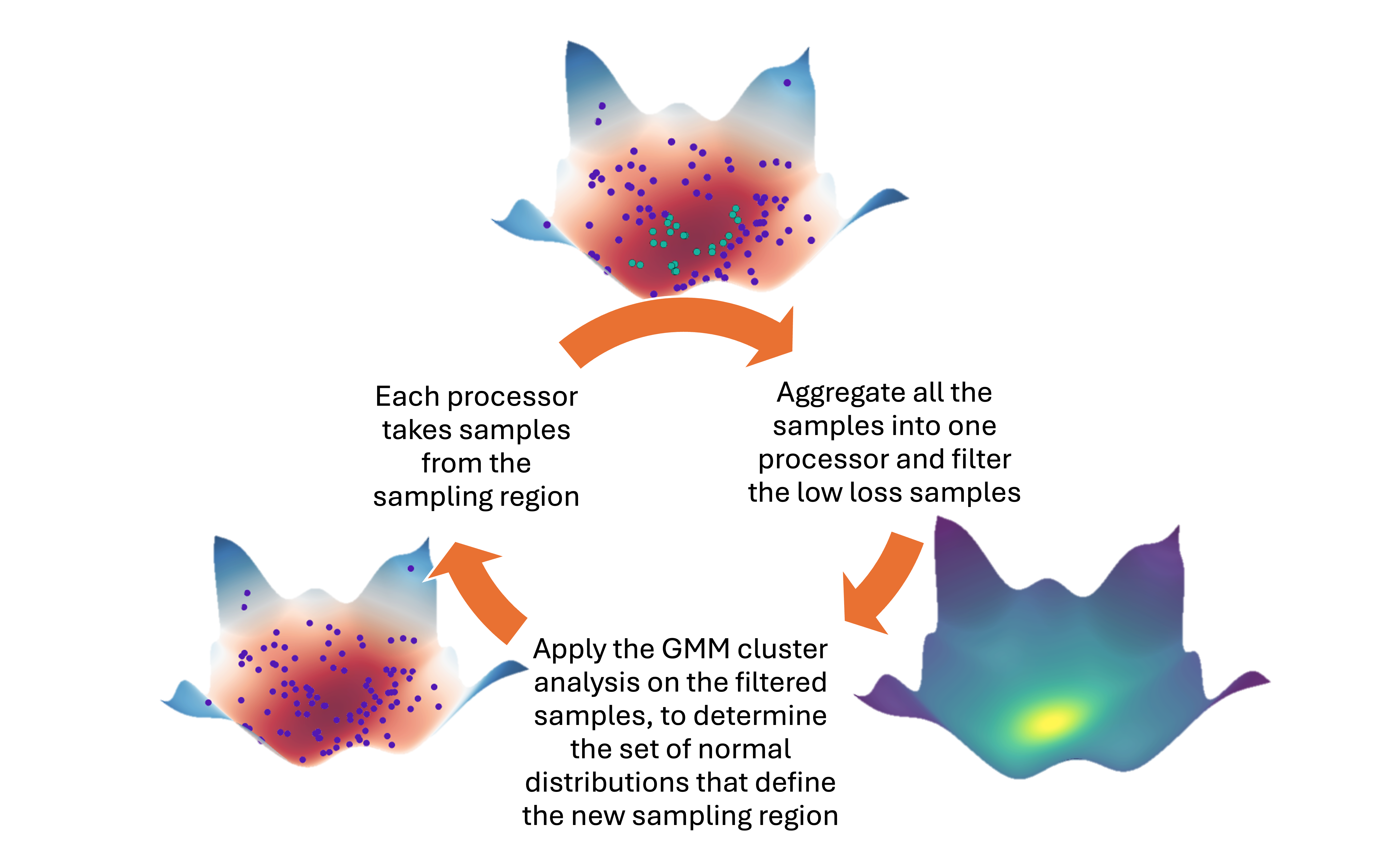}
    \caption{A simple global minimization algorithm, designed to work well for highly parallelizable, non-linear and non-differentiable problems. The algorithm aims to sequentially refine the sampling region by inferring the regions of low loss. After running this algorithm for a few iterations the Simplex method \cite{gao2012implementing} is used to find the minima.}
    \label{fig:fitting_algorithm_full}
\end{figure}

Optimization is non-trivial due to the high dimensionality of $\mathbf{x}$, the non-differentiability of the loss function, and the time-consuming nature of loss evaluations. Various optimization algorithms exist in the literature for such scenarios, including genetic algorithms \cite{mitchell1998introduction} and particle swarm optimization \cite{kennedy1995particle}. However, we found that the algorithm outlined in Figure~\ref{fig:fitting_algorithm_full} provided greater control over the optimization process and generally well converged parameterizations.

The algorithm essentially performs a form of importance sampling, where after each iteration, the sampling region is updated by using a Gaussian Mixture Model (GMM) \cite{reynolds2009gaussian} to form clusters in parameter space regions where the loss is predicted to be low. This algorithm is highly parallelizable, allowing it to be run efficiently on a high performance computing (HPC) machine. Furthermore, it strikes a good balance between exploring the parameter space and converging to a region of low loss. After a set of regions of low loss are found, a local minimizer such as the Simplex method \cite{gao2012implementing} is used to converge to a minimum. We present here the potential that offers a good compromise between accuracy and robustness with respect to the test cases, as shown in section \ref{sc: defect properties}. For efficient exploration of the variable space, we initially fitted sets of parameters independently with respect to each other: First, we fitted the W-He interactions, whilst keeping He-He and H-He to follow DFT vacuum properties. Then, we fine-tuned all of parameters simultaneously to further improve the defect properties whilst maintaining vacuum properties.

\subsection{DFT Data Generation}\label{sc: DFT-calc}

\noindent The DFT data used in this work are primarily sourced from literature. The binding energies of various defect clusters are obtained from Yang \textit{et al.} \cite{yang2018energetics}, and the relaxation volumes of helium clusters from Nguyen-Manh \textit{et al.} \cite{nguyen2015trapping}. However, relaxation volume data for hydrogen-helium clusters are lacking. To address this, we performed DFT calculations using the \textsc{Vasp} \cite{hafner2008ab} program, starting with several unique minimized atomic configurations taken from \textsc{Lammps} \cite{thompson2022lammps}, using an intermediary potential constructed by combing the Mason W-H potential and Juslin's W-He potential. These configurations were further minimized within DFT, from which we obtained defect formation energies and relaxation volumes. All DFT simulations were performed using the GGA-PBE exchange-correlation potential \cite{perdew1996generalized} with cells containing {4\,$\times$\,4\,$\times$\,4} base-centered cubic (bcc) unit cells and a {5\,$\times$\,5\,$\times$\,5} $k$-point mesh. A plane-wave cutoff energy of {450\,eV} was used, with a Methfessel-Paxton \cite{methfessel1989high} electron smearing width of  {0.05\,eV}. The tungsten semi-core shells were treated as valence electrons, resulting in a total number of 12 valence electrons for W, 2 for He, and 1 for H. Forces were converged to {0.01\,eV/\AA}. In both MD and DFT, we computed defect formation energies and relaxation volumes following the method outlined by Ma and Dudarev \cite{ma2019universality}, with the formation energies corrected for periodic elastic interactions using the \textsc{Calanie} \cite{ma2020calanie} code. The resulting formation energies and relaxation volumes are listed in Table~\ref{tab:DFT data table} in the Appendix.


\section{Potential Parameterization}\label{sec:potpar}

\noindent To fit each function in the potential, we begin with a physics-informed prior derived from first-principles studies, followed by a fitting process to further optimize the potential. This approach ensures greater reliability and minimizes the risk of overfitting to the defect dataset \cite{mishin2021machine}.

The Ziegler-Biersack-Littmark (ZBL) stopping model \cite{ziegler1985stopping} serves as the prior in constructing our pair potentials. This alignment is critical for accurately modeling high-energy atomic collisions, where ZBL-based short-range interactions dominate. Equation~\eqref{eq:pair-pot expression} presents the formalism employed, where the fitting process optimizes a set of quintic spline points to minimize the defined loss function.

At short interatomic distances, constraints are applied to enforce adherence to the ZBL stopping model, ensuring correct behavior for small separations. Additionally, as the distance approaches either the cutoff range \(r_\mathrm{c}\) or zero, the potential is designed to transition smoothly to zero, with continuity maintained up to the second derivative. These constraints collectively ensure that the potential is stable both at short  and long ranges:
\begin{equation}
\phi_{\alpha\beta}(r) =
\begin{cases} 
       \mathrm{ZBL}(r) + s_{1}(r) & \text{if } 0 \leq r < r_1 \\
       \mathrm{ZBL}(r) + s_{2}(r) & \text{if } r_1 \leq r < r_2 \\
       \mathrm{ZBL}(r) + s_{3}(r) & \text{if } r_2 \leq r < r_\mathrm{c} \\
       0 & \text{if } r \geq r_\mathrm{c}
\end{cases},
\label{eq:pair-pot expression}
\end{equation}
where we omitted the atomic species indices $\alpha$, $\beta$ in the ZBL and spline functions for brevity.

The general Finnis-Sinclair potential framework permits distinct definitions for electron densities. For instance, tungsten-helium interactions include \( \rho_{\mathrm{W-He}} \), which represents the electron density donated by tungsten to helium, and \( \rho_{\mathrm{He-W}} \), which represents the electron density donated by helium to tungsten. Due to the complex charge localization patterns exhibited by light gases within a lattice, finding a useful prior for these electron densities proved challenging. Therefore the fitting will be the only driving factor in defining the functions. Equation~\eqref{eq:edensity expression} formalizes this approach, with spline constraints ensuring that the electron densities smoothly decay to zero, preserving stability across the interaction range:
\begin{equation}
\rho_{\alpha\beta}(r) =
\begin{cases} 
       s_{1}(r) & \text{if } 0 \leq r < r_1 \\
       s_{2}(r) & \text{if } r_1 \leq r < r_\mathrm{c} \\
       0 & \text{if } r \geq r_\mathrm{c}
\end{cases}.
\label{eq:edensity expression}
\end{equation}

\subsection{Tungsten-Helium Pairwise Potential}

\noindent For greater stability, we incorporated data from quantum mechanical studies \cite{puska1984theory, lang1983interaction} on helium adsorption onto noble metal surfaces. These studies indicate that, unlike hydrogen, helium does not form chemical bonds with metal surfaces, as its interaction energies remain in the meV range. To capture these weak interactions in the tungsten-helium potential, we constrained the W-He pair potential to have a minima close to {3\,\AA} with a dimer energy on the order of a few meV. 


\subsection{Helium Embedding Function}
\noindent The above theoretical studies \cite{puska1984theory, lang1983interaction} have also demonstrated that helium exhibits a repulsive response to electron density, with the degree of repulsion being linearly proportional to the electron density. This behavior has been further corroborated by DFT simulations \cite{ma2021collaborative}, reinforcing the applicability of this functional form.

However, a purely linear form would incorrectly imply that helium could be attracted to regions of negative electron density, leading to erroneous predictions. To avoid this issue, we used a modified functional form:
\begin{equation}
    F_\mathrm{He}(\rho) = \sqrt{a^2 \rho^2 + b^2} - b, \quad \text{where } a \geq 0, b \geq 0.
    \label{eq: embedding expression}
\end{equation}
This ensures physical consistency by preventing attraction to negative electron densities while still capturing the correct repulsive behavior at larger electron densities.

\subsection{Helium-Helium Interactions}

\noindent Traditional He-He pair potentials like the Beck parameterization \cite{beck1968new} are derived from gas-phase experimental data, particularly the second virial coefficient of the virial equation of state \cite{riddell1953theory}. While effective for low-pressure systems up to 2--3\,MPa \cite{DIMIAN2014157}, these potentials prove inadequate for the highly pressurized environments encountered in small helium bubbles in stressed tungsten lattices.

\begin{figure*}[t]
    \includegraphics[width=0.9\textwidth]{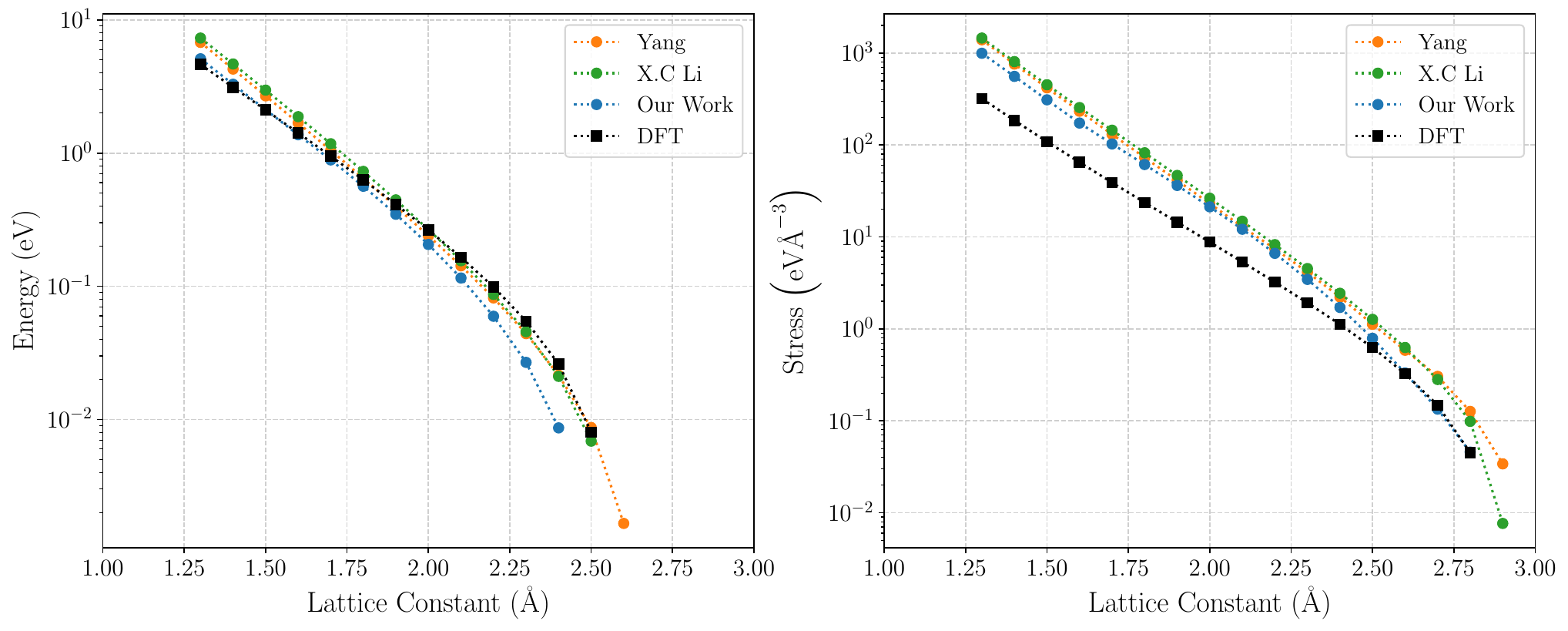}
    \caption{Properties of hcp helium as a function of lattice constant: (a) potential energy and (b) stress (left to right). Our potential exhibits small deviations in the predicted energy of the hcp crystal and accurately captures the low-stress regime. However, at higher stresses, the accuracy decreases as the ZBL term dominates, which is an inherent part of the potential formulation. This trade-off is unavoidable but acceptable. Overall, our potential provides a consistent and reliable representation of hcp helium.}
    \label{fig:hcp_helium}
\end{figure*}

To address this limitation, we developed a revised potential that explicitly incorporates DFT data from the high-pressure regime, where helium adopts a hexagonal close-packed (hcp) structure \cite{mao1988high}. Our fitting strategy balances both the experimental second virial coefficients governing dilute gas behavior, and first-principles energy and stress curves for hcp helium, see Figure~\ref{fig:hcp_helium}. These curves were generated through systematic variation of hcp lattice parameters followed by DFT calculations of total energy and hydrostatic stress. 

The hcp system was initialized with perfect $c/a$ ratio for lattice vectors $(1/2, -\sqrt{3}/2, 0)$, $(1/2, \sqrt{3}/2, 0)$, and $(0, 0, 2 \sqrt{2/3})$, with two He atoms placed at $(0,0,0)$ and $(1/2, 2/3, 1/2)$. We used a {20\,$\times$\,20\,$\times$\,20} $k$-point mesh and a plane-wave cutoff energy of {1000\,eV}. The total energy of the crystal was computed for hcp lattice constants ranging from 1.3 to {4\,\AA}. No structural minimization was performed. The same structures were used to compute the cohesive energies with the interatomic potential.

As shown in Figures~\ref{fig:gas_interactions} and~\ref{fig:hcp_helium}, our potential exhibits good agreement with high-pressure hcp helium data while maintaining reasonable accuracy for low-pressure virial coefficients. By accepting a modest reduction in the accuracy of the second virial coefficient, we achieve a significantly improved description of small interstitial helium clusters within the tungsten lattice. This trade-off does not negatively impact the potential's accuracy for large voids, where helium behaves nearly as an ideal gas. This compromise aligns with our goal in accurately capturing the elastic fields of helium based defects.

During optimization, we found that introducing the many-body interaction as represented by the embedding function provided no statistically significant improvement. To prevent overfitting while maintaining physical realism, we constrained helium’s electron densities to zero, given the minimal contribution of noble gas electrons to the delocalized electron gas. 

\subsection{Hydrogen-Helium Interactions}

\noindent To ensure stability in H-He interactions, we began developing the H-He potential by generating DFT data points for the binding energy of a H-He dimer in vacuum. The binding energies were computed in an orthogonal simulation cell with side lengths of 12, 10, and {10\,\AA}. The H and He dimer was placed along the $x$-direction, and the binding energies were computed for bond distances ranging from 1 to {4\,\AA}. We used a {3\,$\times$\,3\,$\times$\,3} $k$-point mesh and a plane-wave cutoff energy of {1000\,eV}.

Initially, we fitted a pair potential in the form of Equation~\eqref{eq:pair-pot expression} directly to this DFT dataset, omitting any electron density contributions to isolate the pairwise interaction. After achieving a converged fit for the W-He system, we iteratively refined both the electron densities of both H-He and He-H and pair potential parameters with respect to defect and dimer properties. This dual optimization aimed to improve the accuracy of defect properties, e.g. binding energies in tungsten vacancies, while preserving the accuracy of the H-He vacuum interaction derived from DFT.

\begin{figure*}[t]
    \includegraphics[width=0.9\textwidth]{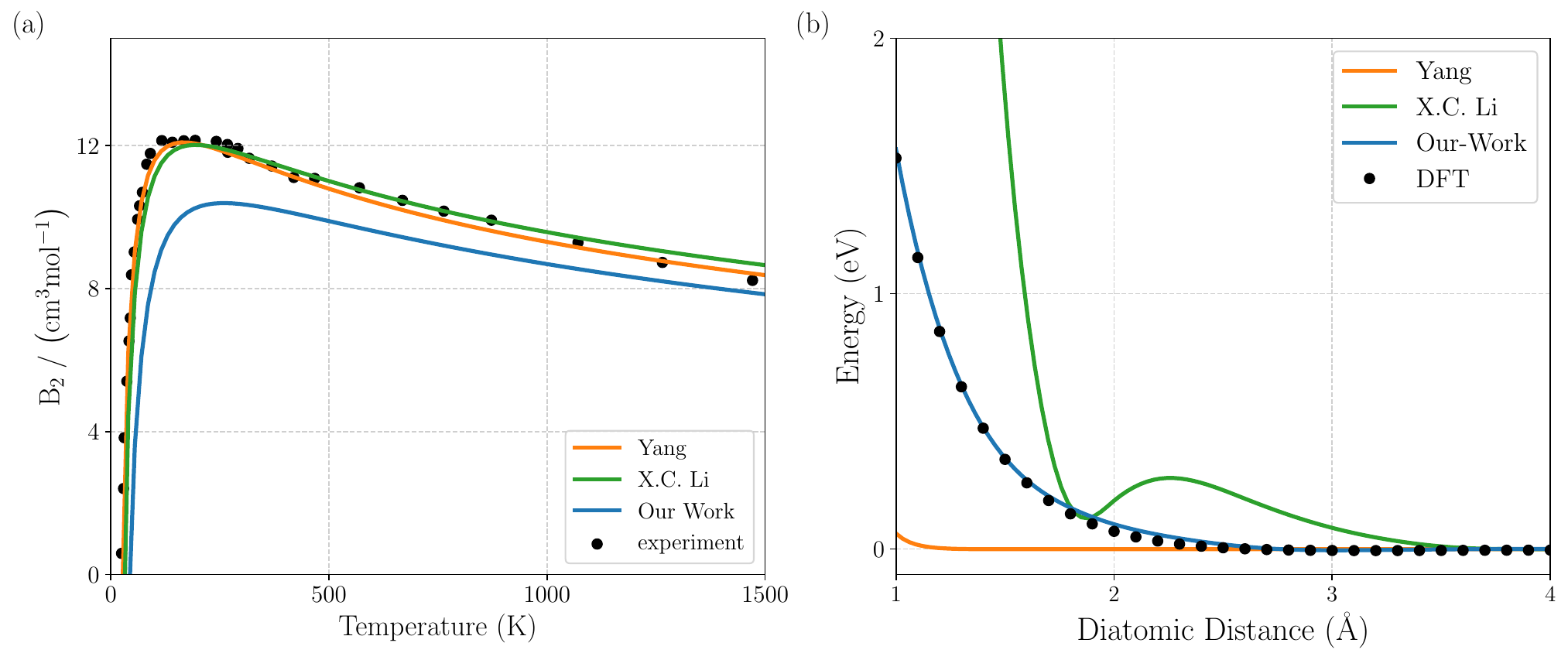}
    \caption{Gaseous properties of helium and hydrogen. (a) The second virial coefficient of helium, which quantifies deviations from the ideal gas law. Experimental data from Beck et al. \cite{beck1968new} is shown, it is to be noted that the Yang potential uses the Beck potential to describe He-He interactions. (b) H-He diatomic interactions, where the Lennard-Jones potential is taken from \cite{belashchenko2006simulation}. This H-He potential has been widely used in the literature \cite{yang2018energetics, bergstrom2017molecular, chen2022md} to model H-He interactions. These plots demonstrate that our potential accurately captures H-He interactions, though minor discrepancies exist in the ideal gas law corrections.}
    \label{fig:gas_interactions}
\end{figure*}

Figure~\ref{fig:gas_interactions} demonstrates that our potential reproduces the DFT-calculated diatomic H-He interaction curve with high precision. This agreement suggests that the energetics of hydrogen and helium within voids or low-density regions will remain reliable. Furthermore, the close adherence to the diatomic interaction indicates that overfitting has been avoided, as the potential’s transferability is validated by its performance in this fundamental limit.

\subsection{DFT Charge Density Comparison}\label{subsec: DFT charge density}
\noindent To assess the qualitative accuracy of our potential, we performed a charge density analysis using DFT calculations. While it is not possible to capture many of the nuances of DFT in an empirical potential optimised for speed, choosing the correct functional forms can significantly enhance the accuracy and reliability of the potential.

Previous potentials \cite{juslin2013interatomic, li2025analytical, bonny2013interatomic} neglect electronic interactions involving helium. This simplification is justified since helium is a noble gas, and therefore has a tightly bound electronic structure. However, in the following charge density analysis we show that there are non-negligible electronic interactions for helium inside a tungsten lattice. 

\begin{figure}[t]
    \subfloat[\label{fig:charge-density-helium-hydrogen}]{
        \includegraphics[width=0.95\columnwidth]{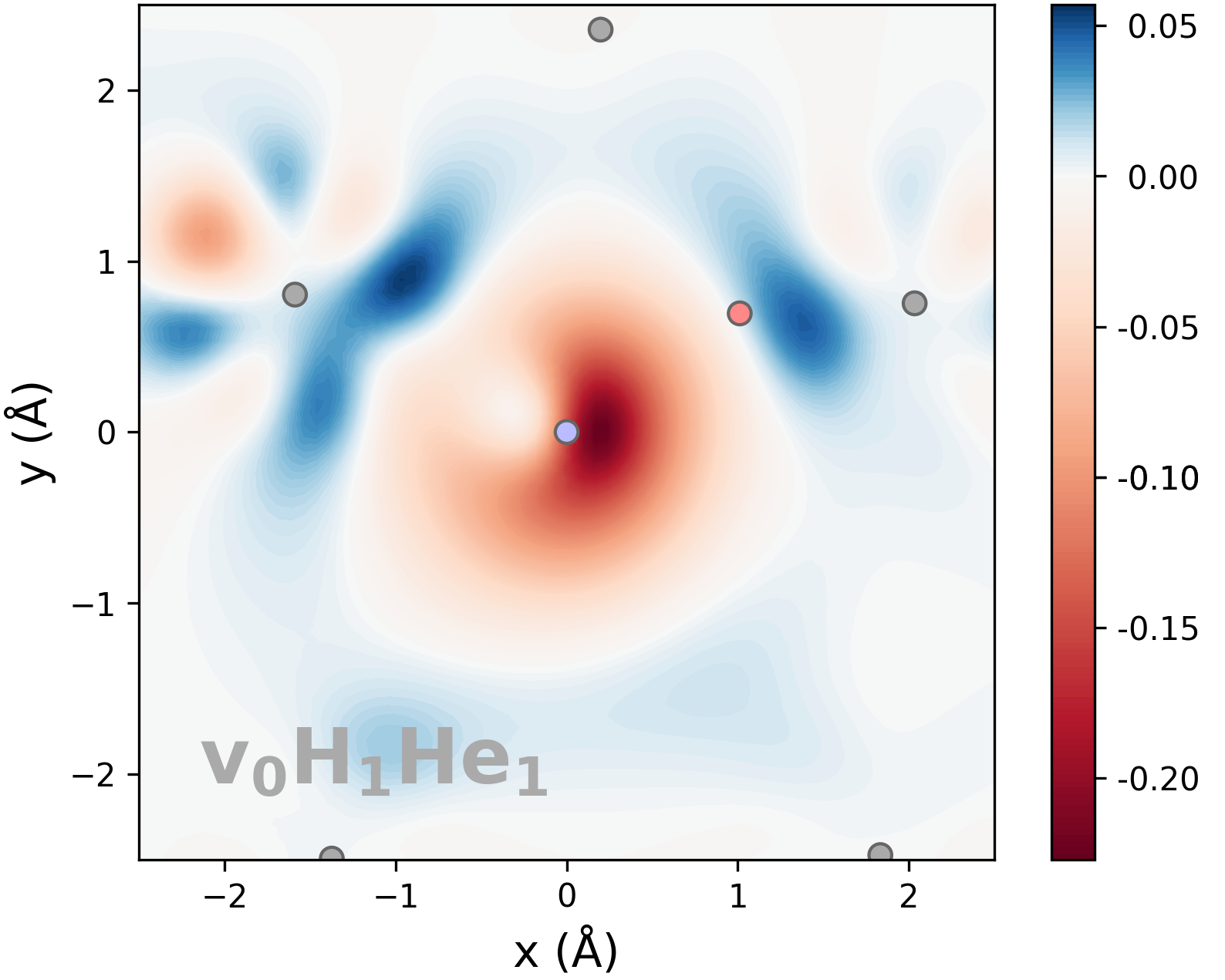}
    }
    \vfill
    \subfloat[\label{fig:charge-density-subsitutional-helium-hydrogen-vacancy}]{
        \includegraphics[width=0.95\columnwidth]{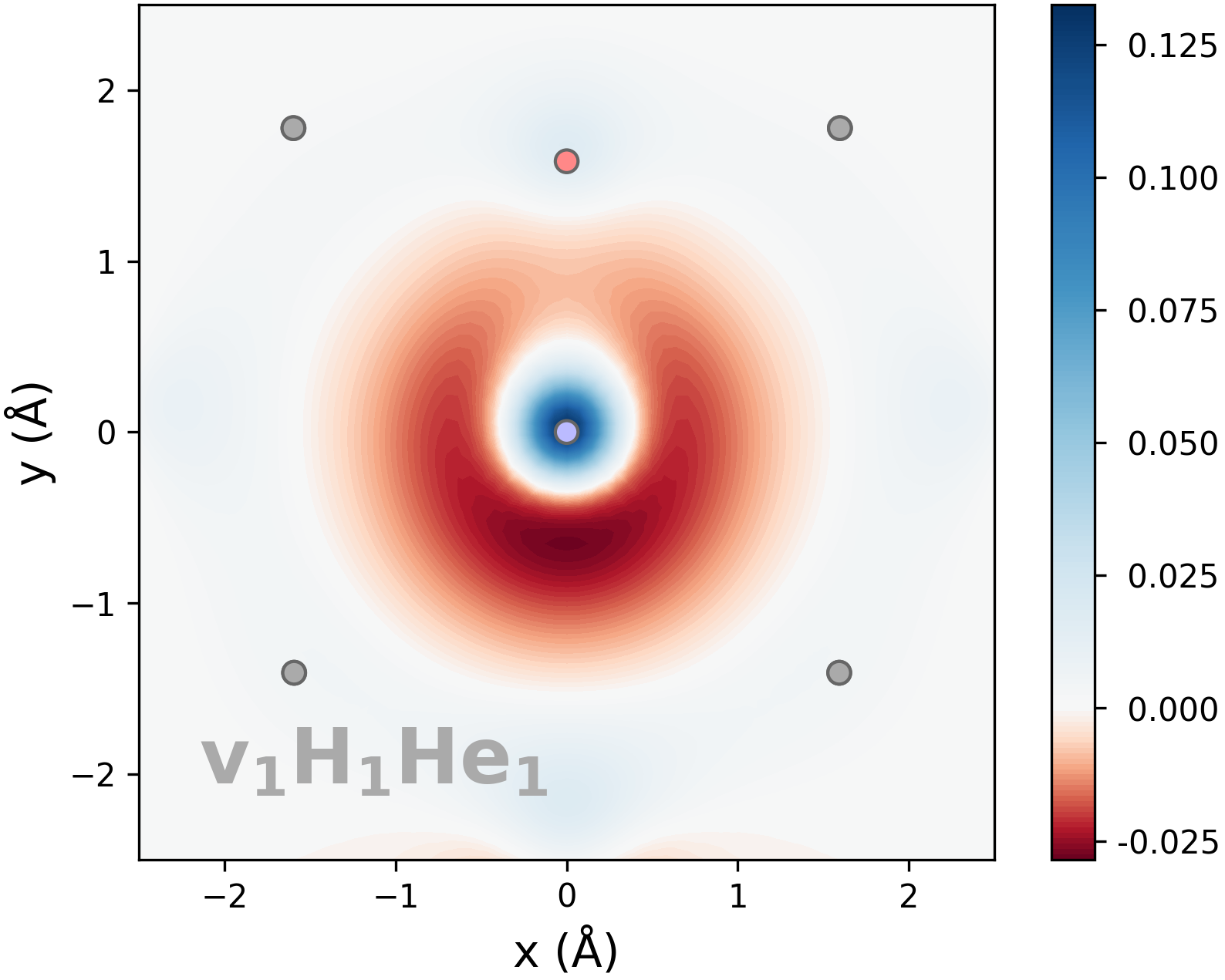}
    }
    \caption{Change in charge density of helium in a H-He interstitial (top) and H-He vacancy (bottom) defect relative to the helium charge density in vacuum, shown here as a slice of the $(100)$ plane intersecting the helium atom. Atoms shown are W (gray), He (light blue), and H (light red).}
    \label{fig:charge density dft plots}
\end{figure}

In Figure\ref{fig:charge density dft plots}, we illustrate how the charge density associated with a helium atom changes when is part of a He-H defect in either an interstitial or vacancy site in tungsten, relative to its state in vacuum. In the interstitial configuration, the charge density increases near the neighboring tungsten atoms and decreases around the helium atom. This trend is consistent with previous findings \cite{wan2018energetics}, which report hybridization between the helium p-orbitals and tungsten d-orbitals, suggesting that there is some a degree of electronic interaction. In the vacancy configuration,the charge density of helium becomes more localized, suggesting that tungsten acts to further confine the helium charge density.

Additionally, we observe a polarization of charge density in the direction of the hydrogen atom, indicating some degree of electronic interaction between hydrogen and helium within the tungsten lattice. Interestingly, the direction of polarization along the He-H axis differs between the interstitial and vacancy configurations. In the interstitial defect, a reduction in charge density is seen around helium, while in the vacancy defect, an increase is observed. This contrast highlights the complexity of electronic interactions in these systems and the importance of including such effects when fitting interatomic potentials.

\section{Defect Properties}
\label{sc: defect properties}
\noindent For the potential to be applicable to highly irradiated microstructures, it must accurately describe the interactions between irradiation-induced defects and light gases. This section examines the potential's accuracy in predicting these interactions.

\subsection{Interstitial Helium Properties}

\noindent We begin by examining the energies and relaxation volumes of a single helium atom within a perfect tungsten lattice. We define the formation energy $E_\mathrm{f}^{D}$ of a defect $D$ as the difference between the total energy of the defect $E(D)$ and the total energies of an equivalent number of constituent atoms in their respective ground states, these being bcc tungsten, diatomic hydrogen in vacuum, and monoatomic helium in vacuum:
\begin{equation}
   E_\mathrm{f}^D  = E(D) - N_\mathrm{W} E_\text{W(bcc)}-\frac{1}{2}N_\mathrm{H} E_\mathrm{H_2} - N_\mathrm{He} E_\mathrm{He}.
\end{equation}

\begin{figure}[t]
    \centering
    \includegraphics[width=0.85\columnwidth]{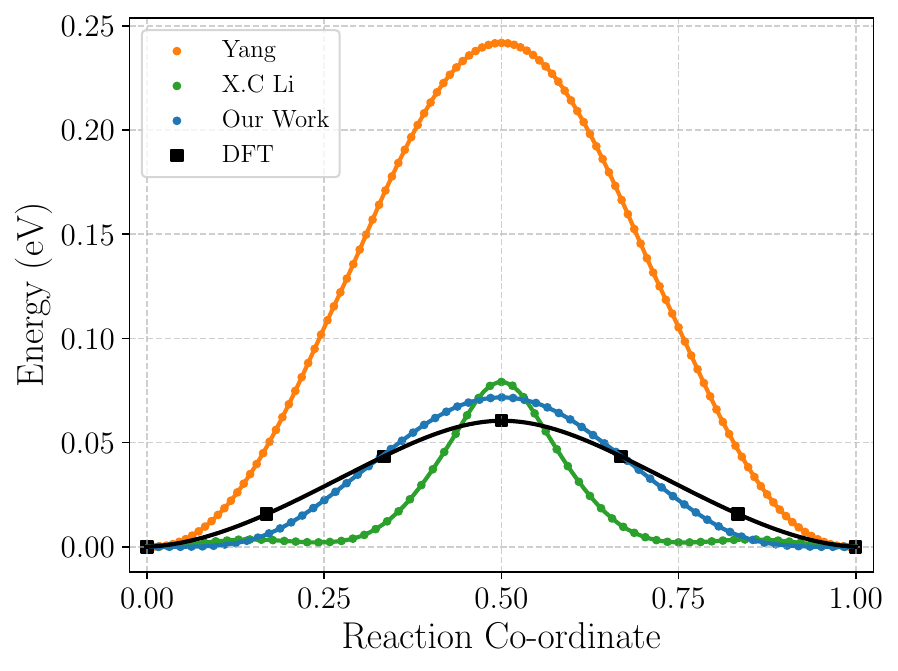}
    \caption{Nudged elastic band (NEB) calculation of the helium migration barrier between two tetrahedral sites. Our potential is in agreement with DFT \cite{tamura2013first}.}
    \label{fig:tet-tet he migration}
\end{figure}

The equilibrium sites of helium within a tungsten lattice dictate diffusion pathways, making their accurate mapping essential. As shown in Table~\ref{tb: helium tungsten lattice}, our potential effectively captures the differences between formation energies in these equilibrium sites. To further analyze kinetics, we compared the minimum energy pathway for the migration of interstitial helium computed by our potential with DFT calculations, as shown in Figure~\ref{fig:tet-tet he migration}. This comparison confirms that our model accurately reproduces the primary energy barrier. However, deviations in the second derivatives along the pathway introduce inaccuracies in zero-point energy estimates, potentially affecting hopping frequency calculations in diffusion studies \cite{vineyard1957frequency}. Despite these discrepancies, the dominant factor in determining hopping frequencies remains the Debye frequency of tungsten atoms, due to their significantly higher mass. Consequently, while minor deviations in diffusion coefficients may occur, they should remain close to expected values.

\begin{table}[t]
\centering
\caption{\raggedright Properties of single helium defects within a perfect tungsten lattice.}
\begin{tabular}{c@{\hskip 0.2cm}c@{\hskip 0.2cm}c@{\hskip 0.2cm}c@{\hskip 0.2cm}c}
 \hline
  \textbf{Property} & \textbf{DFT} & \textbf{Our Work} & \textbf{Yang \cite{yang2018effect}} & \textbf{Li \cite{li2025analytical}}\\
 \hline
  $E_\mathrm{f}^{\mathrm{He, tet}}$ (eV) & 6.16 \cite{becquart2006migration} & \textit{6.73} & \textit{6.68} & 6.21 \\
  $E_\mathrm{m}^{\mathrm{He, tet-oct}}$ (eV) & 0.22 \cite{becquart2006migration} & 0.21 & 0.32 & 0.13 \\  
  $E_\mathrm{m}^{\mathrm{He, tet-tet}}$ (eV) & 0.06 \cite{becquart2006migration} & 0.07 & 0.21 & 0.08 \\
  \(\Omega_\mathrm{rel}^{\mathrm{He - tet}}\) (\(\Omega_0\)) & 0.36 \cite{nguyen2015trapping} & \textit{0.48} & \textit{0.63} & \textit{0.67}\\
 \hline
\end{tabular}
\label{tb: helium tungsten lattice}
\end{table}

Empirical potentials, while widely used in materials simulations, inherently have limitations in accurately reproducing all properties derived from DFT. These limitations necessitate thoughtful compromises. In our study, a discrepancy is the overestimation of the interstitial helium formation energy, denoted as $E_\mathrm{f}^{\mathrm{He, tet}}$. For simulations of bulk tungsten, this discrepancy will only affect the binding energies of helium to large voids, where helium can behave like an ideal gas.  However, for surface simulations, the significance of this property may increase. Nevertheless, due to the high magnitude of $E_\mathrm{f}^{\mathrm{He, tet}}$ and the small relative error, these effects are negligible within typical MD timescales and reasonable temperature ranges.

Another inaccurately predicted property is the relaxation volume of interstitial helium atoms. During the fitting process, we prioritized accurately modeling relaxation volumes for larger defects and those involving vacancies, as these are more prevalent in irradiated materials. Consequently, this compromise introduces inaccuracies in the elastic interactions of interstitial helium atoms with large elastic fields, such as those surrounding dislocations.


\subsection{Helium Interaction with Surfaces}

\noindent Hydrogen binds to tungsten surfaces with an energy of approximately 0.5\,eV \cite{rodriguez2021ab}. In contrast, helium binds much more weakly to metal surfaces \cite{puska1984theory, lang1983interaction}. Our model predicts a weak helium binding energy to the tungsten surface, ranging from 0.05 to 0.06\,eV, depending on surface orientation. While this slightly overestimates the interaction compared to theoretical predictions \cite{puska1984theory, lang1983interaction}, which place helium-surface binding energies on the order of 0.1--{1\,meV}, it remains consistent with the overall trend of weak helium adsorption. Furthermore, such small binding energies are negligible in finite-temperature MD simulations.

\begin{figure}[t]
    \centering
    \includegraphics[width=0.85\columnwidth]{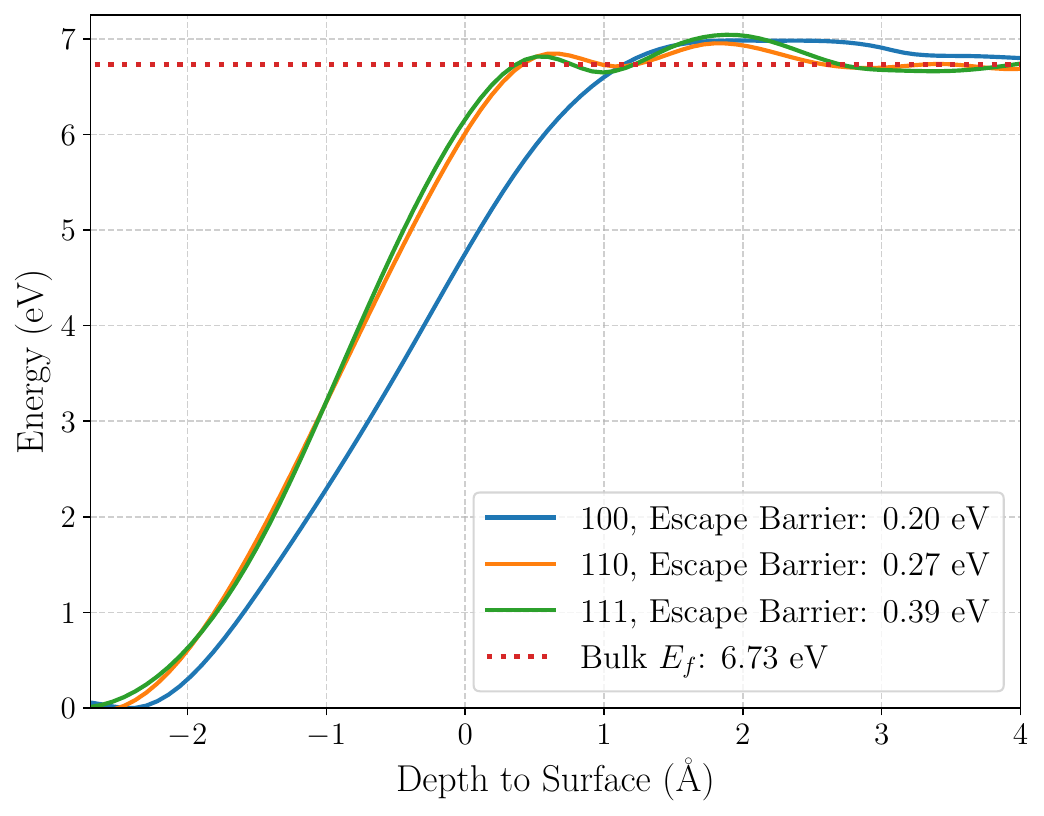}
    \caption{ 
    NEB calculation of helium migration from vacuum into a tungsten surface using our potential. The results reveal small escape barriers, representing the energetic threshold that helium must overcome to enter or escape tungsten surface.}
    \label{fig:surface he migration}
\end{figure}

Wang \textit{et al.} \cite{wang2017embedded} conducted first-principles nudged elastic band (NEB) calculations to study helium migration paths out of surfaces with different orientations. Given the numerous possible migration pathways, these results should be interpreted with caution. Nevertheless, they provide valuable insights into helium-surface interactions. Their calculations indicate significantly higher escape barriers: 0.55\,eV for (100), 0.30\,eV for (110), and 1.03\,eV for (111) surface orientation. As shown in Figure~\ref{fig:surface he migration}, our potential underestimates these barriers but still captures their presence, demonstrating a reasonable extrapolation, since such configurations are not included in the fitting.

\subsection{Energetics of Helium-Hydrogen Clusters}


\noindent For reliable MD simulation, ist is essential to accurately predict the binding energies between combined defects with high accuracy. We define the binding energy for forming a defect \(D_3\) from the constituent defects \(D_1\) and \(D_2\) as:
\begin{equation}
    E_\mathrm{b} = E_\mathrm{f}^\mathrm{D_1} + E_\mathrm{f}^\mathrm{D_2} - E_\mathrm{f}^\mathrm{D_3},
\end{equation}
where a positive \(E_\mathrm{b}\) indicates a favorable reaction $\mathrm{D_1} + \mathrm{D_2} \to \mathrm{D_3}$. 

In this paper, a binding energy will often refer to the energy released when an interstitial gas atom binds to a defect. When multiple light gas atoms are already present in the defect, the number of possible configurations increases exponentially, leading to a distribution of observed binding energies in simulations. To systematically quantify differences between interatomic potentials, we conduct multiple MD simulations at 400\,K for 25\,ps using a simulation cell of 12\,$\times$\,12\,$\times$\,12 unit cells, followed by conjugate gradient (CG) minimization to identify metastable configurations of helium defect clusters. We then retain the minimum energy configurations for further analysis. Clusters that fragmented or underwent trap mutation \cite{boisse2014modeling} were excluded to maintain consistent structural definitions across all potentials. This exclusion criterion ensures direct comparability, as trap mutation, for example, transforms an interstitial helium cluster into a vacancy-helium complex and a self-interstitial, altering the defect's classification.

\begin{figure*}[t]
    \includegraphics[width=0.8\textwidth]{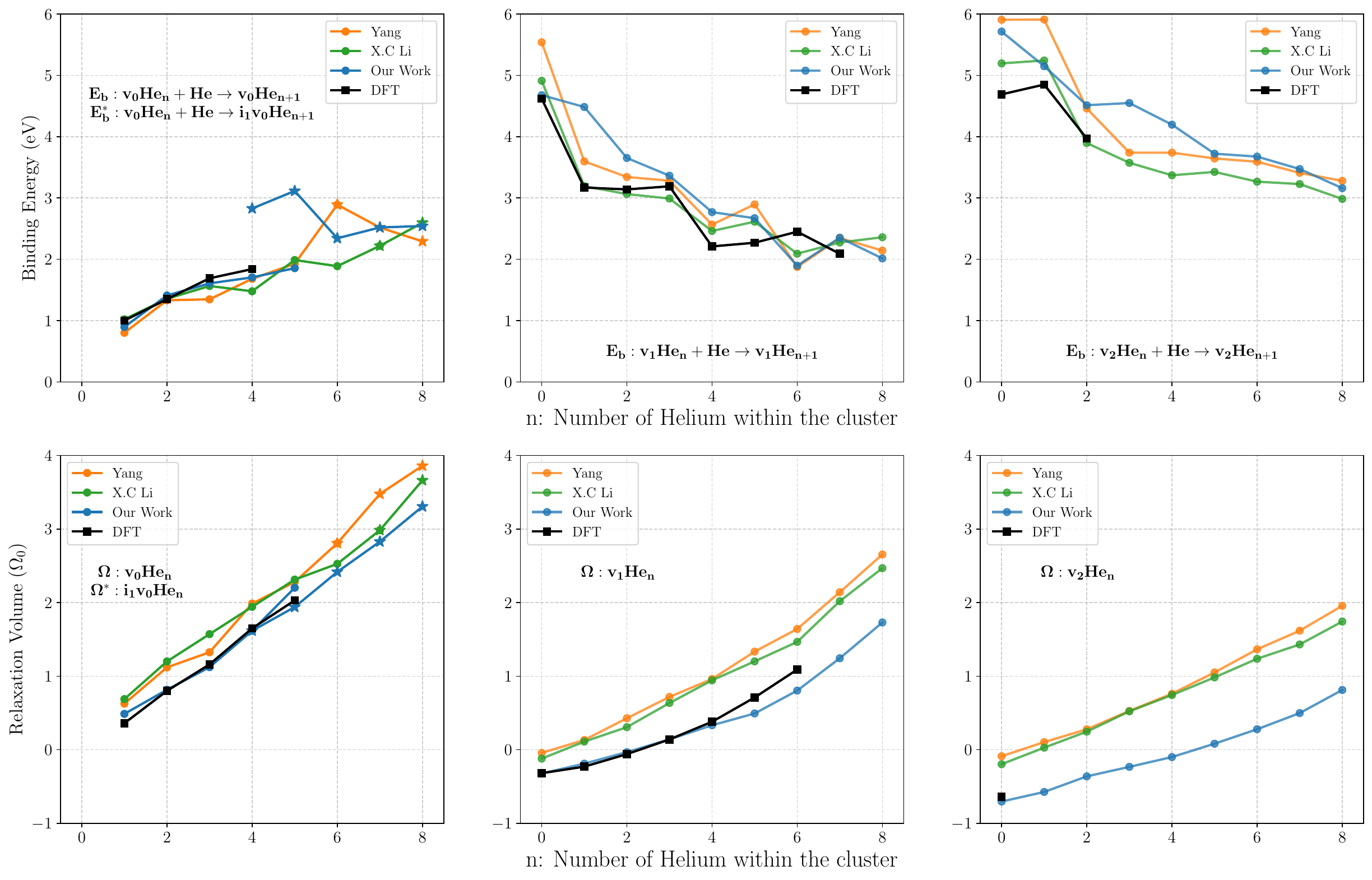}
    \caption{Binding energies (top row) of an interstitial hydrogen atom joining a interstitial helium-hydrogen cluster (a), a vacancy helium-hydrogen cluster (b), and a di-vacancy helium-hydrogen cluster (c). The bottom row shows the relaxation volumes of the corresponding cluster. The 'star' symbol marks self-trapped configurations, where a Frenkel pair was formed.}  
    \label{fig:helium binding}
\end{figure*}

In Figures~\ref{fig:helium binding} to \ref{fig:he_binding_sia}, only binding energies between the found minimum energy configurations are shown. It is to be noted that many metastable configurations exist, and so within a given MD simulation there will be variation in the measured binding energies. Furthermore, we compare our results with DFT calculations from the literature \cite{yang2018energetics, nguyen2015trapping} where available, and perform our own where none exist. Additionally, we compare against other EAM potentials, specifically the {W-H-He} potentials by Li \cite{li2025analytical} and Yang \cite{yang2018effect}.

\subsection{Interactions of Helium with Helium Clusters}
\noindent Figure~\ref{fig:helium binding} demonstrates that all three potentials accurately predict the binding energies of interstitial helium atoms and helium clusters, where trap mutation has not occurred. However, discrepancies emerge in their predictions of relaxation volumes: the Li and Juslin potentials overestimate these volumes compared to DFT-derived data, whereas our potential aligns more closely. This divergence highlights the improved accuracy of our model in capturing the elastic fields of these helium-filled defects, ensuring more realistic representations of defect-induced strain fields, which are critical for understanding helium-driven swelling.

\subsection{Helium driven Trap Mutation}

\noindent Helium driven trap mutation occurs when a large interstitial helium cluster emits a Frenkel pair, forming a vacancy-helium cluster and a self-interstitial atom. This phenomenon, widely reported in atomistic and DFT studies \cite{wilson1981self, you2014clustering, boisse2014modeling, zhang2022effect}, is observed in our simulations and marked by star symbols in Figure~\ref{fig:helium binding} (upper left plot).

Our interatomic potential predicts that trap mutation becomes energetically favorable for helium clusters as small as five atoms. DFT studies report some variation in the critical cluster size required for trap mutation: You \textit{et al.}~\cite{you2014clustering} reported five atoms, Boisse \textit{et al.}~\cite{boisse2014modeling} reported six atoms, and Zhang \textit{et al.}~\cite{zhang2022effect} found a slightly larger threshold of seven atoms. These discrepancies highlight the difficulty of accurately capturing the trap mutation process, likely due to the vast number of possible spatial configurations that must be sampled to definitively identify the minimum-energy state. Given this inherent complexity, we consider our potential’s prediction of a five-atom threshold to be consistent with the range of reported values.


Furthermore, our potential exhibits a pronounced tendency toward trap mutation, stabilizing the resulting configurations by approximately 1\,eV compared to states where trap mutation has not occured. DFT studies \cite{boisse2014modeling, you2014clustering} report that on the onset of trap mutation the energy difference between the two states is quite small. This leads to an overestimation of helium binding energies in practical simulations. While the potential accurately predicts binding energies for clusters of size 5 and 6 in static calculations, the spontaneous onset of trap mutation during dynamic simulations artificially enhances trapping strength. Importantly, this discrepancy does not invalidate the model’s utility: as while qualitatively the model performs correctly however quantitative rate calculations may show discrepancies. An overestimation of trapping energy would only further suppress an already rare event. Moreover, the potential correctly reproduces configurations where trap mutation does not occur, confirming that the overestimation arises from a physical mechanism rather than a systematic error.

The sensitivity of the trap mutation is also demonstrated by considering the variations in the empirical potentials and the DFT data available in literature. Whilst the binding energies of helium to interstitial helium clusters are very similar, their predictions on trap mutation are quite different due to the large configurational space one must explore to find the minimum energy configuration. Therefore it is difficult to quantitatively validate the behavior of trap mutation, while we can still qualitatively under what conditions it occurs.

\begin{figure*}[t]
    \subfloat[Binding energies of interstitial hydrogen joining an interstitial helium-hydrogen cluster (top row), with relaxation volumes of the corresponding clusters (bottom row).\label{fig:hydrogen_interstitial_binding}
    ]{
        \includegraphics[width=0.9\textwidth]{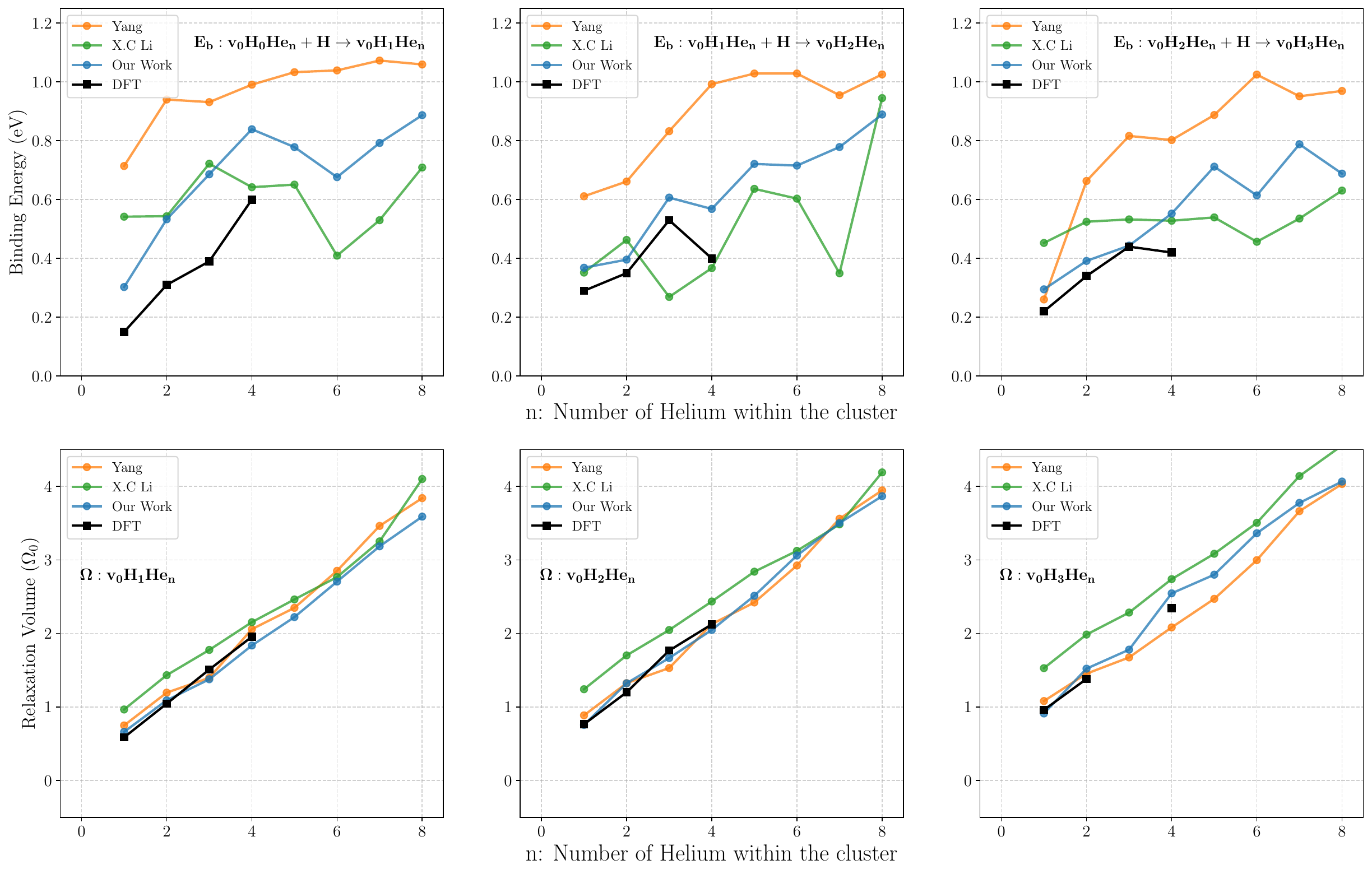}
    }
    \\
    \subfloat[Binding energies of interstitial hydrogen joining an vacancy helium-hydrogen cluster (top row), with relaxation volumes of the corresponding clusters (bottom row).
    \label{fig:hydrogen_vacancy_binding}
    ]{
        \includegraphics[width=0.9\textwidth]{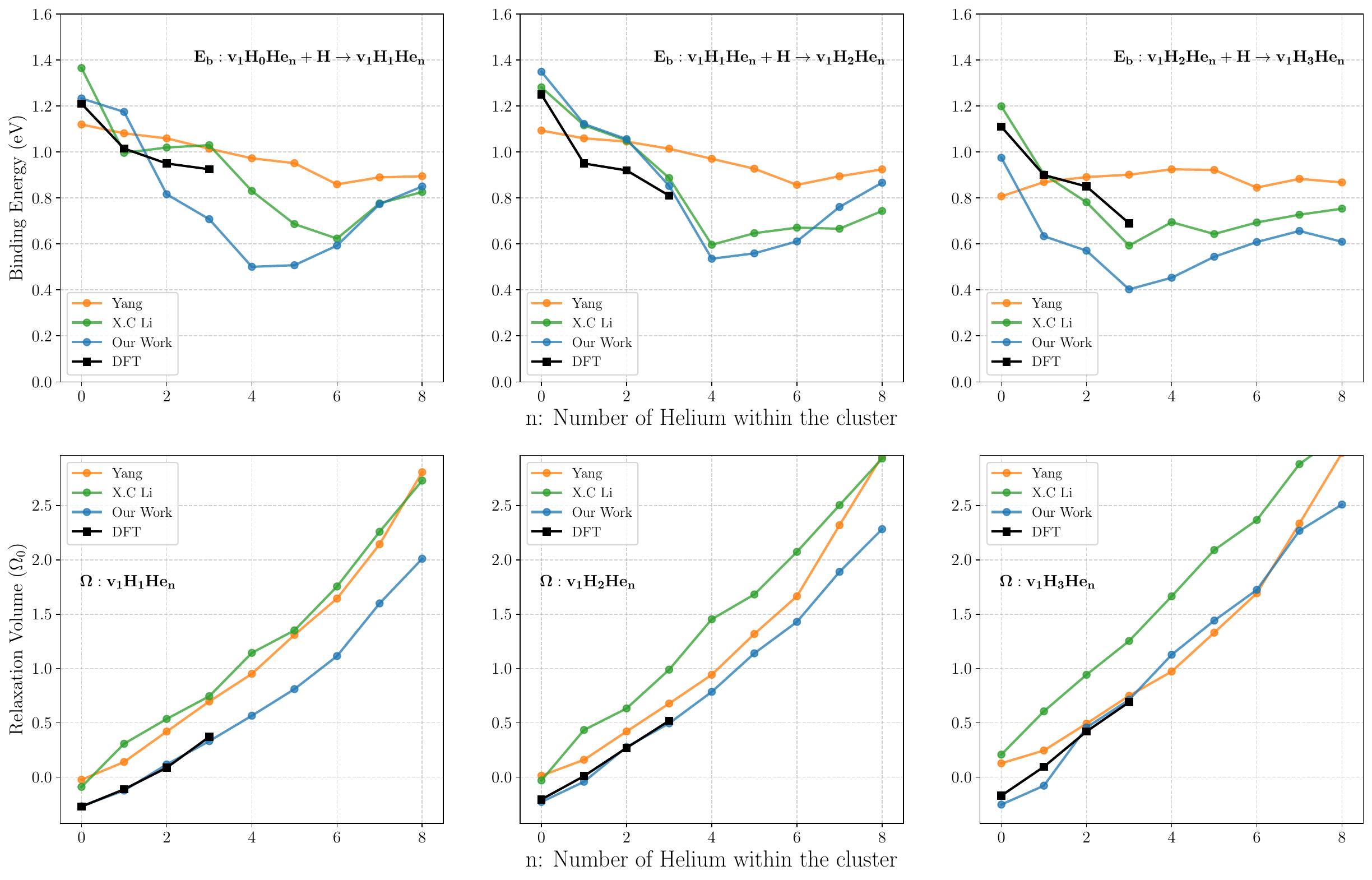}
    }
    \caption{Binding energies of an interstitial hydrogen atom joining an interstitial helium-hydrogen cluster (a) and a vacancy helium-hydrogen cluster (b), including relaxation volumes of the corresponding defects. The initial helium clusters used are the minimum-energy configurations, allowing for self-trapped helium clusters, as these are expected to form in practical simulations.}
    \label{fig:hydrogen_binding}
\end{figure*}


\subsection{Interactions of Hydrogen and Helium Clusters}

\noindent Figure~\ref{fig:hydrogen_interstitial_binding} illustrates a scenario in which helium and hydrogen are implanted into a sample with little to no irradiation, meaning that no vacancies are available as traps. In this case, helium clusters through trap mutation, and DFT results indicate a positive binding between hydrogen and these clusters. Importantly, the energy of this interaction is small, on the order of 0.5\,eV, implying that detrapping occurs regularly even at room temperature. Both our potential and the Li potential accurately capture this weak binding, whereas a significant overestimation is observed with the Yang potential.

In contrast, Figure~\ref{fig:hydrogen_vacancy_binding} depicts a scenario where irradiation results in helium being trapped within vacancies. Here, the binding energy of hydrogen to these clusters is quite significant, on the order of {1\,eV}---similar to the binding energy of hydrogen to an isolated vacancy. Although our potential tends to underestimate this binding for certain configurations, it shows consistent trends with DFT and the Li potential.

Finally, our potential delivers relaxation volumes of these light-gas vacancy clusters in good agreement with DFT.

\subsection{Binding to Interstitial Atoms}

\noindent Self-interstitial atoms (SIA) are the counterparts to vacancies, both of which are generated during irradiation. Unlike vacancies in metals, they are highly mobile \cite{heikinheimo2019direct}, allowing them to migrate easily through the material. When a helium atom approaches an interstitial, the interstitial can shift its position to accommodate the incoming helium atom. As a result, it becomes energetically favorable for helium to bind near these defects, as the interstitial facilitates its integration into the lattice.


%

\begin{figure}[t]
    \includegraphics[width=0.8\columnwidth]{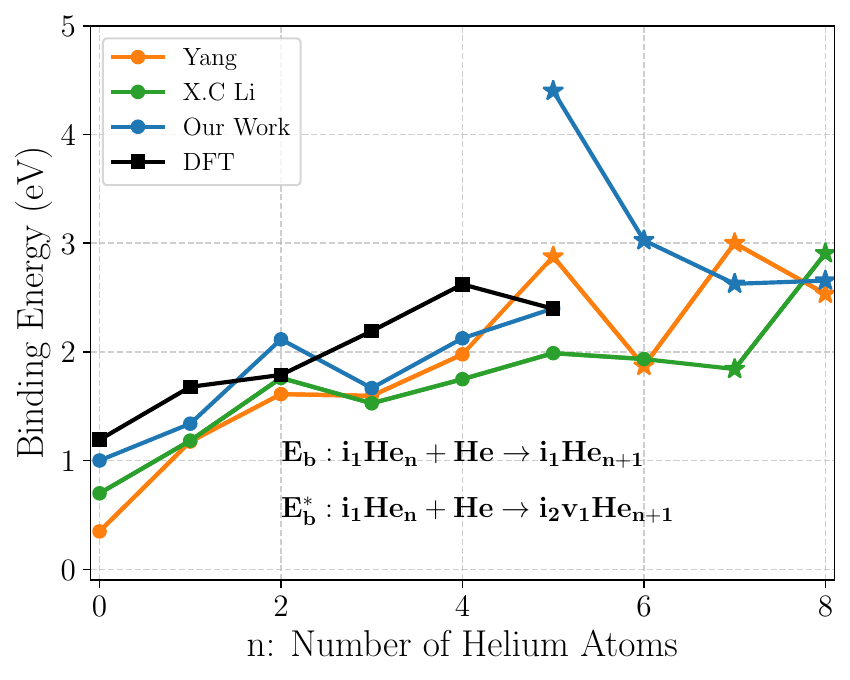}
    \caption{Binding energies of an interstitial helium atom joining a self-interstitial and helium cluster. DFT data from \cite{you2020effects}. Self-trapped configurations are marked with a star ($\star$).}
    \label{fig:he_binding_sia}
\end{figure}

During our fitting process, we specifically considered the binding energy of interstitial helium to a single SIA, as reported in \cite{ma2021collaborative} (1.05\,eV). The binding energies are in good agreement with DFT, as shown in Figure~\ref{fig:he_binding_sia}. We also observe that our potential predicts that a helium cluster of size six undergoes trap mutation, leading to a sharp increase in binding energy. This behavior is expected, as our potential tends to favor trap mutation. 


\subsection{Interactions with Dislocations}
\noindent In the high-dose regime, irradiation-induced defects can accumulate and coalesce, leading to the formation of more complex structures such as dislocation loops, voids, and vacancy loops \cite{ogorodnikova2014annealing, ferroni2015high}. Modeling such large defect structures is typically beyond the practical scope of DFT. To ensure that our potential behaves appropriately when interacting with these defects, we have tested a variety of idealized defect structures using our potential.


\begin{table}[t]
\caption{Binding energy (eV) of helium to various $1/2 \langle 111 \rangle$ dislocation structures in tungsten.}
\begin{tabular}{c@{\hskip 0.2cm}c@{\hskip 0.2cm}c@{\hskip 0.2cm}c@{\hskip 0.2cm}c}
 \hline
  \textbf{dislocation type} & \textbf{DFT} & \textbf{our work} & \textbf{Yang} \cite{juslin2013interatomic} & \textbf{Li} \cite{li2025analytical} \\
 \hline
  perfect edge & 2.96 \cite{bakaev2017trapping} & 2.88 & 3.19 & 3.52 \\
  perfect screw & 1.32 \cite{bakaev2017trapping}&  1.12 & 1.24 & 1.28  \\  
  prismatic loop & - & 1.58 & 1.20 & 2.29 \\
\hline
\end{tabular}
\label{tb: helium tungsten dislocations}
\end{table}

\noindent For both edge and screw dislocations, dislocation core structures were first generated using \textsc{Atomsk} \cite{hirel2015atomsk} based on theoretical elastic displacement fields and subsequently relaxed to an energy minimum using \textsc{Lammps}. The dislocation loop was created by generating a perfect hexagonal loop containing 37 atoms, normal to the $(111)$ plane, and subsequently relaxing the resulting structure. To evaluate binding energies, a helium atom was randomly placed near the dislocation structure, the system was annealed at 500\,K, and then minimized. This was repeated 10 times, and the most stable structure was chosen for the binding energy evaluation. As shown in Table~\ref{tb: helium tungsten dislocations}, the three EAM potentials show variation but on the whole demonstrate good agreement with the limited DFT data available. The results provide confidence in the transferability of this potential to the more complex defect structures occurring under irradiation.




\section{Hydrogen Retention within Helium-filled Voids} 
\noindent Under fusion conditions, voids form in the microstructure, either through vacancy clustering at high temperatures \cite{ipatova2021situ, el2018loop, hasegawa2013neutron} or through the presence of hydrogen and helium, which promote void growth under irradiation \cite{harrison2017study, miyamoto2015systematic}. As shown in the following section, voids exhibit significantly different hydrogen retention behavior compared to an equivalent number of isolated monovacancies.

Both DFT \cite{boisse2014modeling, nguyen2015trapping} and our potential show that helium strongly binds to voids, and through trap mutation, can promote void growth to accommodate additional helium atoms. In contrast, hydrogen exhibits more complex behavior. Hydrogen not only readily adsorbs onto tungsten surfaces \cite{piazza2018saturation, piazza2021predictive}, but it will also form hydrogen molecules within a free volume. In the following, we quantify the binding energy of hydrogen to helium-filled voids using molecular dynamics (MD) simulations with the new potential and present a simple model to explain the observed trends.

\begin{figure*}[ht]
    \includegraphics[width=0.8\textwidth]{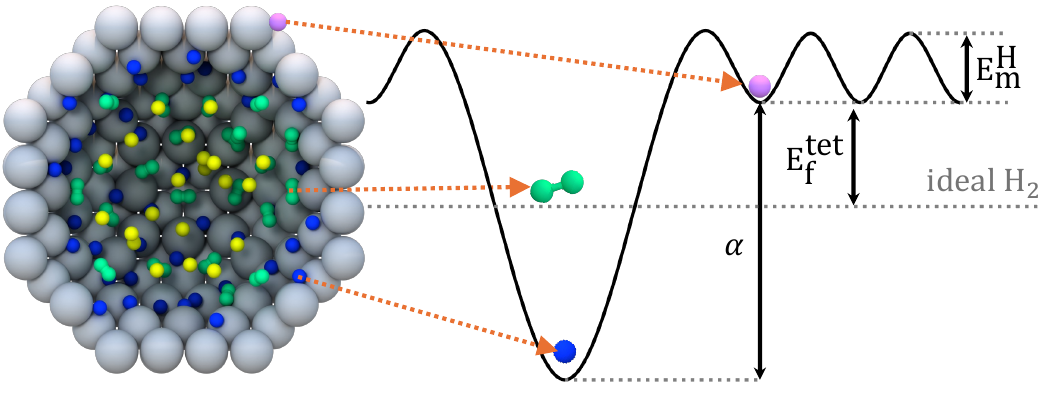}
    \caption{Depiction of the model energy landscape of a hydrogen atom inside and outside of a gas-filled void. The illustration shows the void being filled by surface hydrogen adatoms (blue), diatomic hydrogen molecules (green), and monoatomic helium (white) in its core, as well as hydrogen at interstitial lattice sites (purple). In this model, $\alpha$ represents the binding energy of hydrogen to the surface in the limit  of low occupancy, $E_\mathrm{f}^\mathrm{tet}$ is the formation energy of interstitial hydrogen at the tetragonal lattice site, and $E_\mathrm{m}^\mathrm{H}$ is the migration barrier of interstitial hydrogen.
    }
    \label{fig:void depiction}
\end{figure*}

Figure \ref{fig:void depiction} illustrates the energy landscape of hydrogen within a void. We show that the possible sites a hydrogen atom can occupy within a void in tungsten can be separated into three distinct categories: (i) adatoms bound to the void surface, (ii) molecular hydrogen gas residing within the free volume of the void, and (iii) lattice-bound hydrogen, representing atoms that have escaped into the surrounding tungsten lattice. 
By treating each of these sites separately we arrive at an expression for the total formation energy of a void containing $n_\mathrm{H}$ hydrogen and $n_\mathrm{He}$ helium atoms:
\begin{equation}
    E_\mathrm{f}(n_\mathrm{H}) = E_\mathrm{f}^\mathrm{surf} (n_\mathrm{s}) + E_\mathrm{f}^\mathrm{gas} (n_\mathrm{H_2}, n_\mathrm{He})+ E_\mathrm{f}^\mathrm{lat} (n_\mathrm{L}),
    \label{eq: formation model}
\end{equation}
where $n_\mathrm{s}$, $n_\mathrm{H_2}$, and $n_\mathrm{L}$ denote the number of atoms in the three respective sites, with the total number of hydrogen atoms given by $n_\mathrm{H} = n_\mathrm{s} + 2 n_\mathrm{H_2} + n_\mathrm{L}$. The formation energies are defined in the following sections, $E_\mathrm{f}^\mathrm{surf}(n_\mathrm{s})$ in Eq.~\eqref{eq: surface binding final}, $E_\mathrm{f}^\mathrm{gas}(n_\mathrm{H_2}, n_\mathrm{He})$ in Eq.~\eqref{eq: steric energy}, and $E_\mathrm{f}^\mathrm{lat}(n_\mathrm{L})$ in Eq.~\eqref{eq: lattice model}. Note that this is a highly simplified model---we do not account for free energy, elastic effects, or interactions between the three hydrogen populations. The purpose of the model is to understand the key features of binding in hydrogen-helium filled voids, while the potential itself should be used for high-fidelity binding energy calculations.

Under athermal conditions, the thermodynamic equilibrium corresponds to the configuration that minimizes the total energy for a given hydrogen concentration. Therefore, the occupancy of each site type can be determined by solving the following constrained energy minimization problem:
\begin{equation}  \label{eq:minimization}
    \begin{aligned}
        \min_{n_\mathrm{s},\, n_\mathrm{H_2},\, n_\mathrm{L}} \quad 
        & \left[ E_\mathrm{f}^\mathrm{surf}(n_\mathrm{s}) 
        + E_\mathrm{f}^\mathrm{gas}(n_\mathrm{H_2}, n_\mathrm{He}) 
        + E_\mathrm{f}^\mathrm{lat}(n_\mathrm{L}) \right] \\
        \text{subject to} \quad 
        & n_\mathrm{s} + 2n_\mathrm{H_2} + n_\mathrm{L} = n_\mathrm{H}.
    \end{aligned}
\end{equation}
Furthermore, at finite temperature this model can be extended to describe the free energies of the various populations.

In the following section, we shall introduce and validate models for the formation energies of hydrogen in each of the different sites, leading to a consistent and generalizable model for hydrogen binding to voids in the presence of helium.

\subsection{Model}

We begin with a model of the surface trapping process. As hydrogen binds strongest to the void surface, we expect the void surface to be populated first. The number of the surface trapping sites $N_\mathrm{S}$ is expected to scale with the ratio of surface area to volume of the void. With the void size quantified by the number of constituting vacancies $N_\mathrm{V}$, the resulting expression is
\begin{equation}
    N_\mathrm{S}  = \beta N_\mathrm{V} ^ \frac{2}{3},
    \label{eq: surface site scaling}
\end{equation}
where $\beta$ is a scaling coefficient which is to be fitted to the dataset. Taking the monovacancy as a reference, which can trap up to six hydrogen atoms \cite{heinola2010hydrogen}, we expect $\beta \approx 6$.

The first few hydrogen atoms to bind to the surface will occupy the deepest part of the energy well. As more hydrogen is added and more surface sites become occupied, the repulsive interactions between hydrogen adatoms cause the incremental binding energy to decrease \cite{hou2019predictive, heinola2010hydrogen}. For example, in a monovacancy, the binding energy starts at {1.28\,eV} for the first hydrogen atom and decreases to {0.32\,eV} by the sixth hydrogen. We propose the following scaling law for the surface binding energy: 
\begin{equation}
    E_\mathrm{b}(n_\mathrm{s} | N_\mathrm{S}) = \alpha \left [ 1 -  \left(\frac{n_\mathrm{s}}{N_\mathrm{S}} \right) ^ \gamma\right]
    \label{eq: binding surface}
\end{equation}
where $\alpha$ represents the binding energy of hydrogen in the limit of low occupancy, shown graphically in Figure \ref{fig:void depiction}. 

To find the formation energy, we take note that the incremental binding energy is defined by:
\begin{equation}
    E_\mathrm{b}(n_\mathrm{s}) = E_\mathrm{f}^\mathrm{tet}  + E^\mathrm{surf}_\mathrm{f}(n_\mathrm{s} - 1) -  E^\mathrm{surf}_\mathrm{f}(n_\mathrm{s})
    \label{eq: formation definition}
\end{equation}
and so the formation energy of $n_\mathrm{H}$ hydrogen atoms on the surface is given by:


\begin{equation}
    E_\mathrm{f}^\mathrm{surf} = \left( E_\mathrm{f}^\mathrm{tet} - \alpha \right) n_\mathrm{s} + \alpha \sum_{n=0}^{n_\mathrm{s}} \left(\frac{n}{N_\mathrm{S}}\right)^\gamma
    \label{eq: formation sum}.
\end{equation}
We can approximate the sum using an integral in the limit of large surface occupancies, resulting in:
\begin{equation}
     E_\mathrm{f}^\mathrm{surf} \approx \left( E_\mathrm{f}^\mathrm{tet} - \alpha \right) n_\mathrm{s} +  \frac{\alpha n_\mathrm{s}}{1 + \gamma} \left(\frac{n_\mathrm{s}}{N_\mathrm{S}}\right)^\gamma
     \label{eq: formation integral}.
\end{equation}


With increasing hydrogen content, the low energy surface traps begin saturating and hydrogen will begin forming dimers in the free volume of the void. In our formation energy calculations, the reference state of hydrogen is taken to be an ideal diatomic gas. Consequently, if the hydrogen dimers interact negligibly with the void surface, the formation energy of each dimer is {0\,eV}. However, as the void becomes increasingly filled with hydrogen molecules, intermolecular interactions become significant. At this stage additional energy is required to insert another hydrogen molecule. This repulsive interaction is a steric effect, which must be accounted for under athermal conditions.

In order to account for steric effects, we construct the following model: At {0\,K}, the diatomic gas will settle into its minimum energy configuration. Assuming that the only interaction between molecules is a radial repulsion, the minimum energy configuration would correspond to the one maximizing the packing density of the molecules within the given volume. According to Kepler's conjecture, the packing density is maximized by either face-centered cubic (fcc) or hexagonal close-packed (hcp) structures. X-ray diffraction studies of solid hydrogen suggest an hcp structure \cite{akahama2010evidence}. Therefore, we assume an hcp configuration where only interactions with the first nearest neighbors are significant. Upon this assumption, the total energy of a system of $n_\mathrm{H_2}$ hydrogen molecules in a volume $V$ is given by:
\begin{equation}
    E_\mathrm{f}^\mathrm{gas} = 6 \sum_i^{n_\mathrm{H_2}}
        E \left(\tfrac{1}{2}a_\mathrm{hcp} \right),
    \label{eq: hcp steric energy}
\end{equation}
where
\begin{equation}
a_\mathrm{hcp} = \left (\frac{V}{3 \sqrt{2}  n_\mathrm{H_2}} \right)^\frac{1}{3}.
\end{equation}

Assuming a Lennard-Jones (LJ) type repulsion of the form
\begin{equation}
E(r) \sim r^{-12},
\end{equation}
the energy can be expressed simply by:
\begin{equation}
    E_\mathrm{f}^\mathrm{gas} =k \left( \frac{n_\mathrm{H_2}}{V}\right) ^ 4n_\mathrm{H_2} 
    \label{eq: steric energy simple}
\end{equation}
where $k$ is simply a constant dependent on the gaseous interaction.

As hydrogen binds to the void surface, a thin shell adjacent to the surface is not accessible to molecular hydrogen. This is corroborated by Hou et al. \cite{hou2019predictive} who show that molecular hydrogen was not found within small vacancy clusters $N_\mathrm{V} < 3$. Hence introducing an exclusion volume to our model is crucial to prevent excessive formation of hydrogen molecules. To account for this, we first approximate the total void volume as the atomic volume of tungsten multiplied by the number of vacancies, then determine the equivalent spherical radius. By subtracting an exclusion distance $\delta r$ from this radius, we obtain the following expression:
\begin{equation}
    V = \frac{4 \pi }{3} \left [a \left( \frac{3 N _\mathrm{V}}{8 \pi}\right)^\frac{1}{3} - \delta r\right]^3
    \label{eq: void volume}.
\end{equation}

The effect of helium gas on the formation energy is also treated in the steric energy model. We adapt the steric energy to account for the H-He and He-He interactions by the following expression:
\begin{equation}
    E_\mathrm{f}^\mathrm{gas} =  \frac{\left(n_\mathrm{H_2} + n_\mathrm{He}\right)^3}{V^4}
    \left( k_\mathrm{1} n_\mathrm{H_2} ^ 2 + 
    2 k_\mathrm{2} n_\mathrm{H_2}n_\mathrm{He} + 
    k_\mathrm{3}n_\mathrm{He}^2 \right)
    \label{eq: steric energy}
\end{equation}
where $k_1, k_2, k_3$ denote the LJ parameters for each of the {H$_2$-H$_2$},{H$_2$-He} and {He-He} interactions respectively.

The interstitial hydrogen atoms are all assumed to have the same formation energy $E_\mathrm{f}^\mathrm{tet}$, since the population of available interstitial lattice sites is much greater than the number of hydrogen in the lattice; interstitial hydrogen is treated as non-interacting. The total formation energy of the lattice hydrogen is hence given by:
\begin{equation}
    E_\mathrm{f}^\mathrm{lat} = E_\mathrm{f}^\mathrm{tet} n_\mathrm{L}
    \label{eq: lattice model}
\end{equation}

By combining the formation energies in equations \eqref{eq: formation sum}, \eqref{eq: steric energy}, and  \eqref{eq: lattice model}, we can express the total formation energy of the void, see Equation~\eqref{eq: formation model}. Given a concentration of hydrogen and helium, we can then find the formation energy and the occupancies of the various sites by solving the constrained minimization problem \eqref{eq:minimization} numerically. To find the binding energy of a hydrogen to a void, we can take the finite difference between the formation energies as given by Equation~\eqref{eq: formation definition}.

\subsection{Parameter Inference - Steric Effects}

The steric parameters are the simplest to infer. To determine them, we construct a 5×5×5 hcp supercell, containing 500 lattice sites, with varying lattice constants. Each site is initialized with either a hydrogen molecule ($\mathrm{H_2}$) or a helium atom, according to the composition of the mixture. After initialization, the system energy is minimized, and the resulting formation energy is computed by subtracting the energy contribution of the $\mathrm{H_2}$ bonds. Finally, the steric parameters $k_1, k_2, k_3$ are obtained through a fit, minimizing the least-squares difference between the MD-simulated and model energies. The uncertainties are calculated by estimating the covariance matrix from the Hessian of the loss function at the minima \cite{pyomo_covariance}.

\begin{figure}[t]
    \includegraphics[width=0.8\columnwidth]{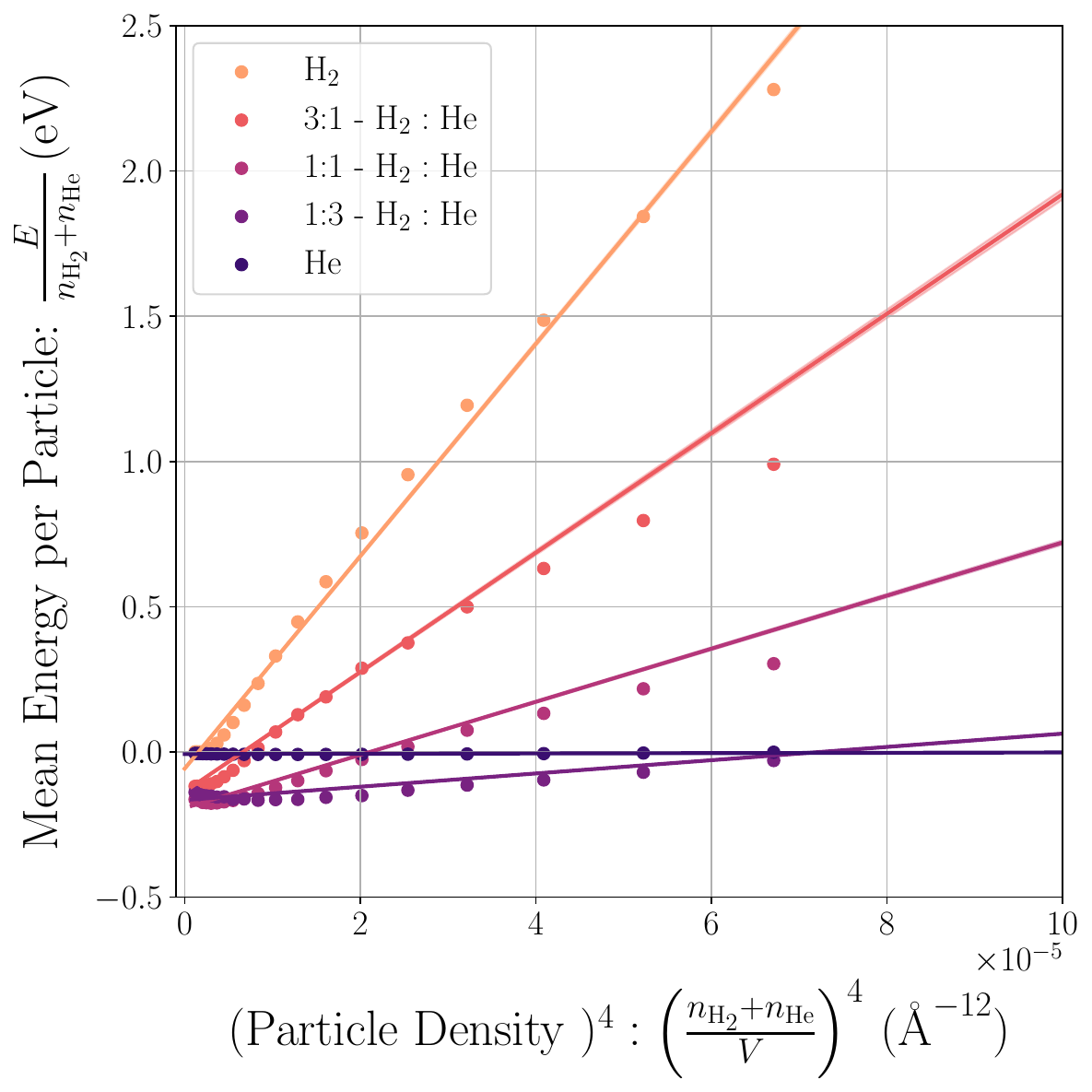}
    \caption{Steric energy of a minimized hcp structure containing hydrogen molecules and helium atoms. Shown here is the energy per gas atom over the fourth power of the total particle density. Pure $\mathrm{H_2}$ (orange) exhibits the highest steric energy, while increasing the helium fraction progressively lowers the steric energy. }
    \label{fig:steric_energy}
\end{figure}

\begin{table}[t!]
\centering
\setlength{\tabcolsep}{10pt}      
\caption{Lennard–Jones interaction parameters for the various gas-gas interactions}
\begin{tabular}{c l c}
\toprule
parameter   & interaction        & value (eV\,\AA$^{12}$) \\
\midrule
$k_1$ & $\mathrm{H_2-H_2}$ &  36500 ± 140 \\
$k_2$ & $\mathrm{H_2-He}$  &  0.0 \\
$k_3$ & $\mathrm{He-He}$   &  58.6 ± 0.1\\
\bottomrule
\label{tb:steric_parameters}
\end{tabular}
\end{table}

Figure \ref{fig:steric_energy} illustrates the quality of the fit and table \ref{tb:steric_parameters} lists the resulting fitted parameters. The steric energy is clearly dominated by $\mathrm{H_2-H_2}$ interactions, which is roughly two orders of magnitude larger than the corresponding $\mathrm{He-He}$ interactions. The fit shows slight deviations at lower particle densities and higher helium concentrations, likely due to attractive contributions that are not included in our model. Additionally, the fit indicates that $\mathrm{H_2-He}$ interactions contribute negligibly to the overall repulsion. This behavior can be attributed both to the unmodeled attractive component—evidenced by the negative intercept in Figure \ref{fig:steric_energy}—and to the repulsive contribution being effectively captured by the $\mathrm{H_2-H_2}$ interactions.

\subsection{MD Simulation of Voids}

To determine the remaining model parameters, we perform MD simulations to obtain formation energies of light-gas-filled voids. The simulations are initialized with a $15 \times 15 \times 15$ supercell at the equilibrium lattice parameter at {0\,K}. A void is created by successively deleting atoms with high potential energy. A set concentration of hydrogen and helium atoms is then introduced to the void, and the system is annealed from {400\,K} to {200\,K} over {250\,ps} using a Langevin thermostat with a damping time constant of {100\,ps}. Finally, the energy of the system is minimized using the method of conjugated gradients to obtain a local energy minimum. For each given hydrogen and helium concentration, we perform 32 simulations that only differ in the random seed, and take the lowest energy configuration as our estimate of the global energy minimum.  

To approach the global energy minimum more efficiently, we initialize the gas inside the void following a schema informed by our model. As an initial guess, we set $\alpha = 1.28$ eV, $\beta = 6$ and $\gamma = 2$, values fitted to the monovacancy binding energies, and assume $\delta r = 0$. Using these parameters, we estimate the occupancies ($\theta_X = n_{X}/n_\mathrm{V}$) of the various populations and initialize the gas atoms accordingly. By annealing the system and performing multiple (32) independent simulations, we aim to enhance sampling and improve the chances of reaching a good estimate of the global minimum.  


\subsection{Parameter Inference -- Surface Parameters}

We begin by examining Equation \eqref{eq: binding surface}, which introduces a power-law scaling for the incremental surface binding energy. This, in turn, implies via Equation \eqref{eq: formation integral} that the total formation energy also follows a power-law dependence.  A straightforward way to test this hypothesis is to plot
\[
y \equiv \frac{E_\mathrm{f}}{n_\mathrm{H}} 
\quad \text{against} \quad 
x \equiv \frac{1}{1 + \gamma}\left(\frac{n_\mathrm{H}}{N_\mathrm{V}^\frac{2}{3}}\right)^{\gamma}.
\]

\begin{figure}[htbp]
    \centering
    \includegraphics[width=0.93\columnwidth]{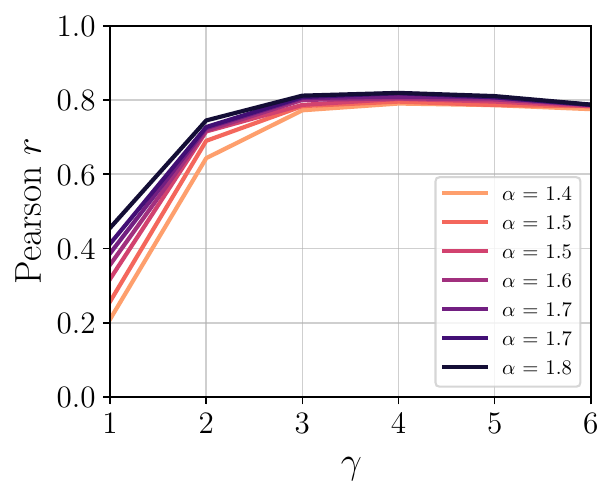}
    \includegraphics[width=0.93\columnwidth]{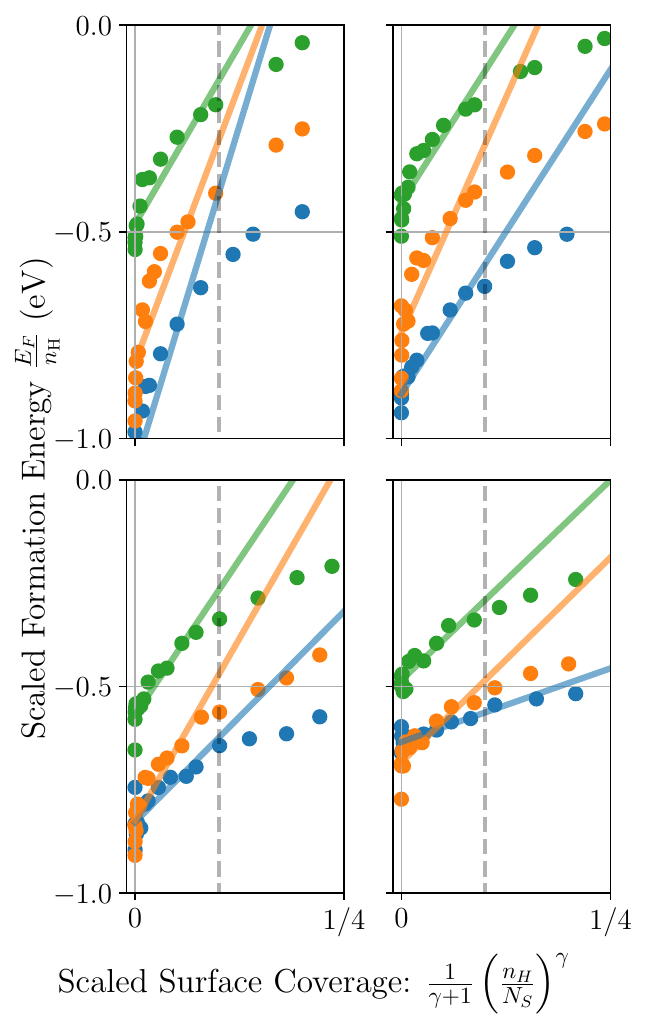}
    \caption{
        (Top) Mean Pearson \(r\)-value averaged over different vacancy sizes, helium contents, and all trial \(\beta\) values for each \(\alpha\). The \(r\)-value is maximized at \(\gamma = 4\), which was therefore selected.  
        (Bottom) Normalized formation energy as a function of surface occupancy in transformed coordinates for different void sizes. Markers indicate MD data points, while the solid lines show the fits, see Eq.~\eqref{eq:scaled-affine}, up to the transition hydrogen content, indicated by the gray dotted line. Colors correspond to helium occupancy: blue, orange, and green represent \(\theta_\mathrm{He} = 0\), \(\theta_\mathrm{He} = 0.5\), and \(\theta_\mathrm{He} = 1\), respectively. These plots use \(\gamma = 4\) applied to the full dataset. A trial value of \(\alpha = 1.6\) and \(\beta = 6\) was used to appropriately scale the plots for visualization.
        }
    
    \label{fig:hydrogen void binding transformed}
\end{figure}

If hydrogen atoms occupy only the surface sites, the power law reduces to an affine relation,
\begin{equation}\label{eq:scaled-affine}
    y = \frac{\alpha}{\beta^\gamma} x + \bigl(E_\mathrm{f}^\mathrm{tet} - \alpha\bigr),
\end{equation}
with slope $\alpha\beta^{-\gamma}$ and $y$-intercept $E_\mathrm{f}^\mathrm{tet} - \alpha$.


The model also provides a criterion for the onset of molecular hydrogen formation: this occurs when the hydrogen binding energy at the surface equals $E_\mathrm{f}^\mathrm{tet}$. The corresponding hydrogen content at this transition point is then defined as $n_\mathrm{H}^\mathrm{T}$:
\begin{equation}
    \left ( \frac{n_\mathrm{H}^\mathrm{T}}{N_\mathrm{V}^\frac{2}{3}} \right)^\gamma= \beta ^ \gamma\left( 1 - \frac{E_\mathrm{f}^\mathrm{tet}}{\alpha} \right).
\end{equation}

In this representation, $\gamma$ is chosen such that the MD data points collapse onto a straight line for $n_\mathrm{H} < n_\mathrm{H}^\mathrm{T}$, i.e., up to the onset of molecular hydrogen formation. Accordingly, we select the value that maximizes the Pearson correlation coefficient $r$ uptill the transition point $n_\mathrm{H}^\mathrm{T}$. The exponent $\gamma$ is restricted to natural numbers to allow for an analytical expression of the surface formation energy. Since the transition point $n_\mathrm{H}^\mathrm{T}$ is not yet known, we test several values of $\alpha$ (1.4--{1.8\,eV}) and $\beta$ (4--8) and then choose the $\gamma$ that performs the best overall.

Figure \ref{fig:hydrogen void binding transformed} shows that $\gamma = 4$ provides a good fit to the dataset, with the data points showing strong linear correlation below the transition hydrogen content. Using $\gamma = 4$ for the surface formation energy \eqref{eq: formation sum} results in the following expression:
\begin{equation}
\begin{aligned}
    E_\mathrm{f}^\mathrm{surf} =\;& (E_\mathrm{f}^\mathrm{tet} - \alpha) n_\mathrm{s} +\\
    & \frac{\alpha}{30 N_\mathrm{S}^4} \big[
        n_\mathrm{s} (1 + n_\mathrm{s}) (1 + 2n_\mathrm{s})
    (3n_\mathrm{s}^2 + 3n_\mathrm{s} - 1)
    \big]
    \label{eq: surface binding final}
\end{aligned}
\end{equation}

\begin{table}[t!]
\centering
\setlength{\tabcolsep}{10pt} 
\caption{
    Variation in \( \alpha \) and \( \beta \) averaged over void size for 3 different helium occupancies, inferred by fitting a straight line to the transformed MD data, for \( \gamma = 4 \).
}
\label{tb:linear_surface_parameters}
\begin{tabular}{c c c c}
\toprule
Parameter & \( \theta_\mathrm{He} = 0 \) & \( \theta_\mathrm{He} = 0.5 \) & \( \theta_\mathrm{He} = 1 \) \\
\midrule
\( \alpha \) (eV) & \( 1.6 \pm 0.2 \) & \( 1.5 \pm 0.1 \) & \( 1.3 \pm 0.1 \) \\
\( \beta \)       & \( 5.5 \pm 1.0 \) & \( 5.0 \pm 1.0 \) & \( 5.0 \pm 0.5 \) \\
\bottomrule
\end{tabular}
\end{table}


Table \ref{tb:linear_surface_parameters} shows that void size scatters these parameters with respect to void size, with greater variation with respect to helium content. So it is still important to address the cause of these fluctuations.

The variations in $\beta$ can be attributed to two factors. First, vacancy clusters are not generally perfect spheres, particularly when they are small. For example, a void consisting of 16 vacancies corresponds to an approximately spherical cluster with an additional monovacancy attached. As a result, the surface area does not scale smoothly with the surface-to-volume ratio, leading to discrete jumps in the number of available sites at small cluster sizes and thereby introducing some scatter. 

The second source of variation in $\beta$ arises from an artifact of the potential, which exhibits an attraction between the void surface and hydrogen molecules. Consequently, newly formed hydrogen molecules adopt energies similar to those of surface adatoms, making it appear as though more sites are available than expected. We note that these observations are made at {0\,K}; at finite temperature, thermal effects are expected to smooth out these artifacts. Since we lack a clear physical motivation for scaling $\beta$ with void size, and because such scaling would imply that the number of sites does not follow the expected surface-area to volume ratio ($2/3$ power), we simply adopt a constant $\beta$ to ensure predictability and to prevent overfitting.




The variation of $\alpha$ is most pronounced for smaller voids, whereas larger voids show only minor discrepancies. The dependence of $\alpha$ (the binding energy of hydrogen to the void at low occupancies) has been reported in both MD and DFT studies \cite{mason2023empirical, wang2017embedded, hou2019predictive}. Modeling this behavior is non-trivial, partly due to the highly non-linear interaction between hydrogen and tungsten, which involves both embedding and pairwise effects, and partly because void surface sites have distinct binding energies depending on the local surface topology \cite{hou2019predictive}. To ensure transferability, we therefore adopt a constant value of $\alpha$ across all vacancy sizes, even though this choice may not be perfectly accurate for every individual cluster.

The second source of variation in $\alpha$ arises from the helium occupancy. Owing to the comparatively simpler interaction between hydrogen and helium, it is possible to construct a physically motivated relation to capture this effect. Referring back to gas energy in Equation \eqref{eq: hcp steric energy}, we note that the average separation between gas atoms scales with the inversely cube root of the particle density. Consequently, the distance between an introduced hydrogen atom and a helium atom will also follow this dependence.   

Using the pair potential fitted in Figure \ref{fig:hcp_helium}, we apply a power-law approximation to the repulsive interaction, which is found to scale as $r^{-4}$. Combining these results, the repulsive energy between an introduced hydrogen atom and the helium gas within the void can be expressed as:
\begin{equation}
    \delta E \propto \left(\frac{N_\mathrm{He}}{V} \right) ^ \frac{4}{3},
\end{equation}
where the volume $V$ is given by Equation \eqref{eq: void volume}. Assuming that this repulsive energy is simply added to the binding energy, we find:
\begin{equation}
    \alpha = \alpha_0 -  \lambda \left(\frac{N_\mathrm{He}}{V} \right) ^ \frac{4}{3},
    \label{eq: alpha corrected}
\end{equation}
where $\alpha_0$ is the binding energy of hydrogen to the void surface in the limit of low hydrogen occupancy and in the absence of helium in the void.

Equation \eqref{eq: alpha corrected} also accounts for the non-linearity observed in Figure \ref{fig:hydrogen void binding transformed}, where helium exerts a strong influence on small voids but a much weaker effect on larger ones. This behavior is governed by the parameter $\delta r$ which excludes a thin spherical shell of volume from the total available volume, see Equation \eqref{eq: void volume}. At small void sizes, this exclusion represents a significant fraction of the total volume, whereas for larger voids the effect becomes negligible.

\begin{figure}[htbp]
    \centering
    \includegraphics[width=1.0\columnwidth]{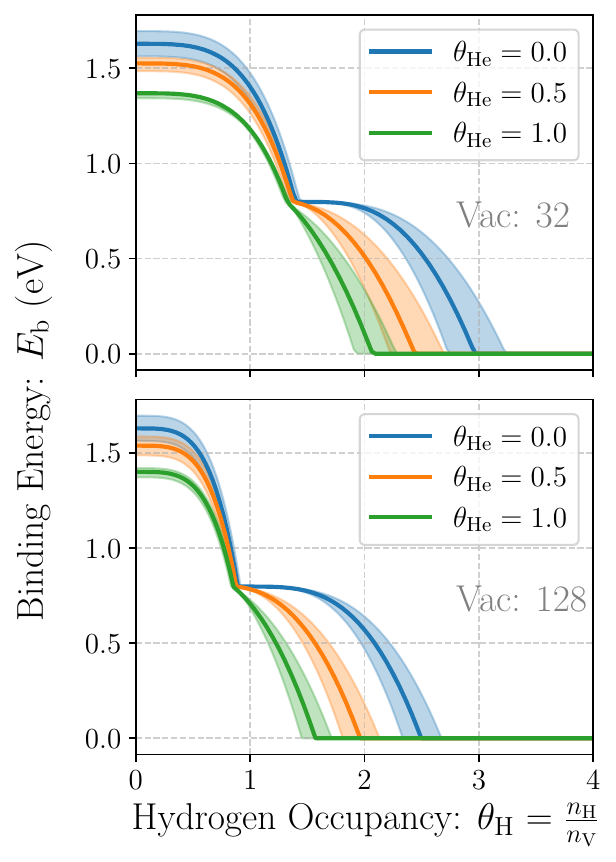}
    \caption{Plot of predicted binding energy against the hydrogen $\theta_\mathrm{H}$ occupancy for three different helium occupancies $\theta_\mathrm{He}$, using the 'full fit' parameters described in Table \ref{tb:fitted_parameters}, for void sizes 32 and 128. The two discontinuities indicate formation of molecular hydrogen, and subsequently hydrogen escaping into the surrounding tungsten lattice.
    }
    \label{fig:hydrogen void binding fitted}
\end{figure}

\begin{table}[t]
\centering
\setlength{\tabcolsep}{6pt} 
\caption{Optimized parameters obtained from fitting our model to the MD dataset. Reported uncertainties correspond to one standard error. We give two sets of parameters, one fitted to the entire dataset ('full fit'), and one fitted to only the larger voids $N_\mathrm{V} > 8$ ('large void fit')}
\label{tb:fitted_parameters}
\begin{tabular}{c c c c}
\toprule
parameter & full fit &  large void fit & units \\
\midrule
$\alpha_0$   & $1.63 \pm 0.01$   & $1.55 \pm 0.01$   &eV \\
$\beta$    & $5.4 \pm 0.05$   & $6.2 \pm 0.1$    &-- \\
$\delta_r$  & $0.40 \pm 0.03$   & $0.50 \pm 0.05$   & \AA \\
$\lambda$ & $7.2 \pm 0.4$   & $5.5  \pm 0.3$     & eV\,\AA$^{4}$ \\
\bottomrule
\end{tabular}
\end{table}

\begin{figure}[htbp]
    \centering
    \includegraphics[width=1.0\columnwidth]{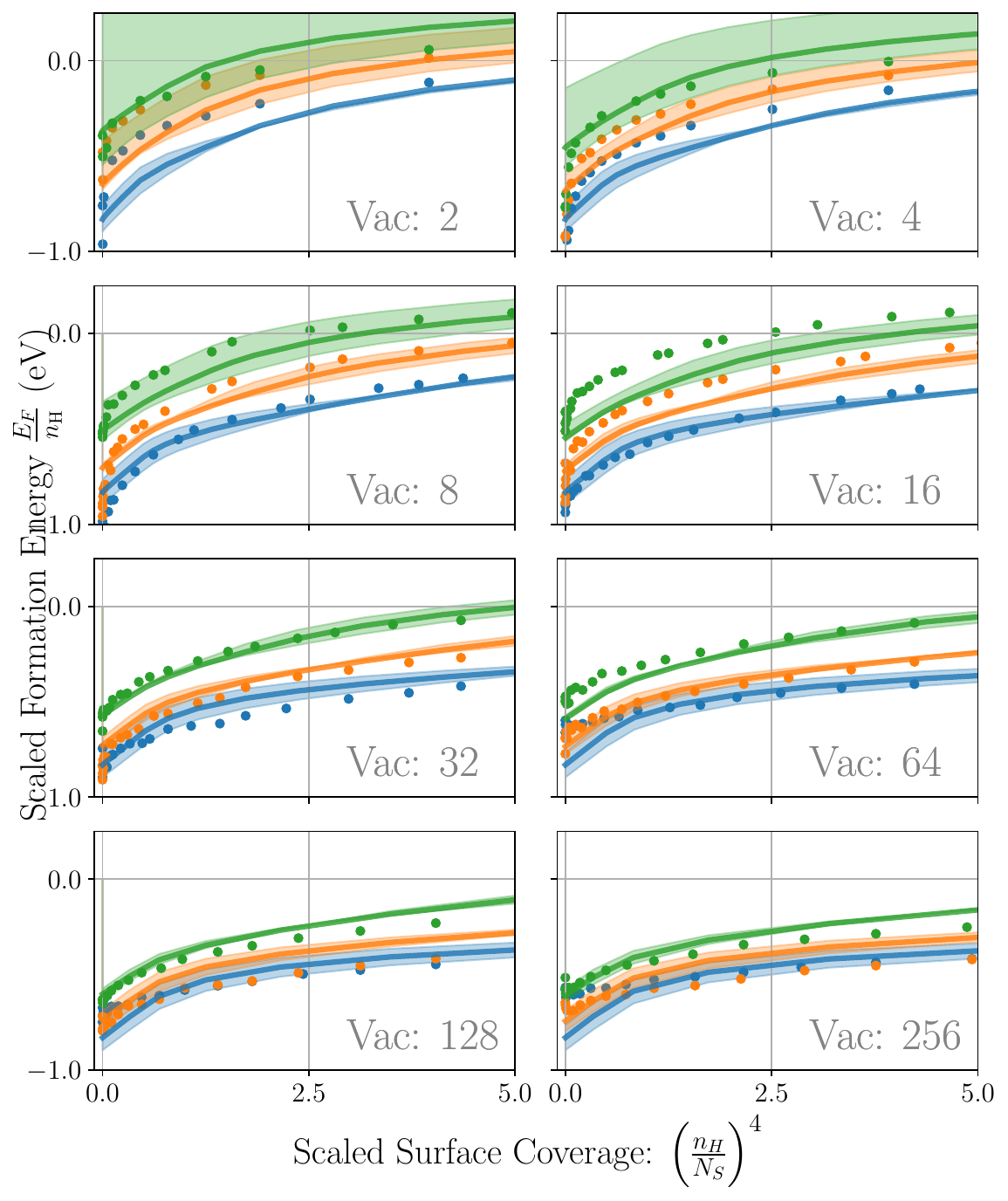}
    \caption{Plot of normalized formation energy with respect to the surface occupancy to the fourth power, the colours blue, green and orange represent helium occupancies ($\theta_\mathrm{He}$) of 0, 0.5, and 1, respectively. The solid markers represents the MD datapoints while the curves represent the model prediction for the 'full fit' parameters described in Table \ref{tb:fitted_parameters}.
    }
    \label{fig:hydrogen void transformed fitted}
\end{figure}

Due to the variability of $\alpha$ and $\beta$, we performed two separate fits: one for the entire dataset and one focused on the larger voids, which exhibit much less variability, see Table~\ref{tb:fitted_parameters}). We recommend using the 'full fit' parameters when the void size distribution is unknown, as they perform well across all void sizes. However, if the void sizes are known and predominantly large, the 'large void fit' parameters are preferred, as they are tailored to larger voids but may not provide optimal predictions for smaller ones.

The fitting procedure is straightforward: the loss function was defined as the sum of squared errors of the scaled formation energies ($E_\mathrm{F}/n_\mathrm{H}$) and subsequently minimized with the Simplex algorithm \cite{gao2012implementing}. To estimate the standard errors of the fitted parameters, we employed a Hessian-based approach. The Hessian of the loss function was evaluated at the minima and used to construct the covariance matrix, from which the standard errors of the parameters were obtained \cite{pyomo_covariance}.

Figure \ref{fig:hydrogen void transformed fitted} illustrates the quality of our fit for the 'Full Fit' parameters given in Table \ref{tb:fitted_parameters}. The model reproduces most void sizes and helium occupancies very well, but deviations appear for the smaller vacancy clusters. These discrepancies stem from the discrete jumps in the number of sites and the variation of $\alpha$. In addition, the fit is especially sensitive to the exclusion distance $\delta r$ at the smallest void sizes. This sensitivity arises because removing a constant radial thickness eliminates a large fraction of the total void volume, and the resulting nonlinear effect amplifies the impact of $\delta r$.

Figure \ref{fig:hydrogen void binding fitted} shows the binding energies predicted by our model for 32-vacancy and 128-vacancy voids at three different helium occupancies. The two observed discontinuities correspond to changes in the filling mechanism: the first discontinuity indicates the formation of molecular hydrogen, while the second marks the escape of hydrogen from the void, signaling saturation. 

Another notable feature is the influence of helium. Beyond the first discontinuity, the presence of helium rapidly increases the steric energy of the gas within the void, leading to much faster saturation. We add a cautionary note regarding the use of our model at high helium occupancies: Equation \eqref{eq: alpha corrected} was derived under the assumption of a simple helium–hydrogen interaction, neglecting helium–tungsten interactions. At high helium concentrations, however, helium is expected to interact with the void surface. Therefore, we consider the model to be applicable for helium occupancies of up to $\theta_\mathrm{He} \sim 1$, which we expect to lie well in the range of what is encountered under fusion conditions.

From the perspective of hydrogen retention modeling, it is evident that larger voids retain proportionately less hydrogen. As the void size increases, the relative number of surface sites becomes vanishingly small, and more hydrogen can potentially be stored as diatomic gas in a bubble.

The maximum number of hydrogen atoms that can be accommodated in a void is reached once the surface sites are occupied and the energy required to form molecular hydrogen exceeds the formation energy of interstitial hydrogen. In the limit of large voids, this condition can be expressed as
\begin{equation}
\frac{\partial E_\mathrm{f}^\mathrm{gas}}{\partial n_{\mathrm{H}_2}} = E_\mathrm{f}^\mathrm{tet}.
\end{equation}
Substituting the gas formation energy \eqref{eq: steric energy simple} and the definition of the void volume in terms of tungsten atomic volumes yields:
\begin{equation}
\frac{n_\mathrm{H_2}}{N_\mathrm{V}} = \frac{a_0^3}{2}\left( \frac{E_\mathrm{f}^\mathrm{tet}}{5 k_1}\right)^\frac{1}{4}.
\end{equation}
To arrive at the maximum occupancy, we consider that each H$_2$ molecule contains two hydrogen atoms and also include the number of hydrogen atoms on the fully occupied void surface, see Equation~\eqref{eq: surface site scaling}, resulting in the simple expression for the hydrogen occupancy at saturation:
\begin{equation}
\theta_\mathrm{H}^\mathrm{sat} 
= \beta N_\mathrm{V}^{-1/3}
  + a_0^3 \left(\frac{E_\mathrm{f}^\mathrm{tet}}{5k_1}\right)^{1/4},
\end{equation}
which in the limit of an infinitely sized void yields
\begin{equation}
\theta_\mathrm{H}^\mathrm{sat}(N_\mathrm{V} \to \infty) \approx 1.4.
\end{equation}
A similar analysis can be carried out for a helium-filled void, see Equation \ref{eq: steric energy}, however, owing to the quartic form of the expression, the resulting solution is not easily expressed in closed form.

We observe that this value is three times smaller than the typically predicted maximum retention for a monovacancy. This highlights the strong size dependence of hydrogen retention and underscores that accurate models must account for the void size distribution.

\section{Conclusion and Outlook}

\noindent In this study, we have developed an empirical potential for the  W-H-He system specifically for reproducing ab initio binding energies and relaxation volumes of irradiation-induced defects. We have demonstrated its robustness across a wide range of defect types, including vacancies and dislocations. A key contribution of this work is the explicit fitting of relaxation volumes, an aspect often overlooked in prior studies that focused primarily on defect energetics. Relaxation volumes are critical for defining the elastic fields associated with defects and play a significant role in predicting microstructural properties such as eigenstrains.

While we found the potential to perform well across the investigated properties, we would like to note some limitations. These include an overestimation of the energy released during helium-driven trap mutation, an underestimation of the second virial coefficient of helium gas, and a slight overestimation of the formation energy of helium in tungsten. The elevated energy favors trap mutated configurations; however, as the critical cluster size for trap mutation is predicted consistently with DFT, this is not expected to affect most applications significantly. The underestimated second virial coefficient introduces minor inaccuracies in helium gas behavior, though this is likely negligible due to helium's near-ideal behavior. The overestimated helium formation energy is unlikely to impact most molecular dynamics simulations, though it may affect predictions involving helium detrapping from the bulk to the surface or to voids.

In the final section of this study, we present a thermodynamically motivated model to capture the energetics of light-gas-filled voids. 
Our interatomic potential is unique in its ability to replicate tungsten surface energies whilst providing a reasonably accurate description of hydrogen–tungsten interactions, making it well-suited for these predictions. We demonstrate the applicability of our minimal model across a range of void sizes, from the smallest vacancy cluster, the di-vacancy, to large voids of size 256, and for varying helium concentrations. The model is based on a physically motivated description of a void containing hydrogen and helium, providing confidence in its predictive capability and transferability. Moreover, the simplicity of the model makes it well-suited for integration into larger, component-scale hydrogen retention simulations.

For future work, our potential can be used to model gas-filled voids at finite temperatures, including all free energy contributions. In a fusion reactor, large temperature gradients and irradiation damage leads to void formation, and understanding the transport of hydrogen through these gradients is crucial. This work provides a key component for such comprehensive modeling.

\acknowledgements

\noindent We thank Xiao-Chun Li for providing the \textsc{Lammps} potential file to their W-H-He potential we used in our comparison. This work has been carried out within the framework of the EUROfusion Consortium, funded by the European Union via the Euratom Research and Training Programme (Grant Agreement No 101052200 -- EUROfusion) and from the EPSRC [grant number EP/W006839/1]. To obtain further information on the data and models underlying this paper please contact PublicationsManager@ukaea.uk. Views and opinions expressed are however those of the authors only and do not necessarily reflect those of the European Union or the European Commission. Neither the European Union nor the European Commission can be held responsible for them. The authors acknowledge the use of the Cambridge Service for Data Driven Discovery (CSD3) and associated support services provided by the University of Cambridge Research Computing Services (\url{www.csd3.cam.ac.uk}) in the completion of this work.


\appendix

\counterwithin{figure}{section}
\counterwithin{table}{section}

\section{Pair Potential Parameters}

\noindent The pair potentials are mathematically expressed by Equation~\eqref{eq:pair-pot expression}, and the fitted parameters are a set of knot points with corresponding function values, derivatives and second derivatives. Each pair potential is described by of three quintic splines, parameterized by four sets of knot points. The spline knot parameters are listed in Tables~\ref{tb:whe params} to \ref{tb:hhe pair params}.

\begin{table}[ht]
    \caption{Quintic spline knot parameters for tungsten-helium pair potential $\phi_{\text{W}-\text{He}}$}
    \begin{tabular}{c@{\hskip 0.5cm}c@{\hskip 0.5cm}c@{\hskip 0.5cm}c}
        \hline
        \textbf{$r$} & \textbf{$f(r)$} & \textbf{$f'(r)$} & \textbf{$f''(r)$} \\
        \hline
        0.00000000 & 0.00000000 & 0.00000000 & 0.00000000 \\
        1.50385307 & -1.46343606 & 4.08015088 & 2.21413655 \\
        2.17164089 & -0.42714108 & 0.41350907 & -0.03906926 \\
        4.85138892 & 0.00000000 & 0.00000000 & 0.00000000 \\
        \hline
    \end{tabular}
    \label{tb:whe params}
\end{table}

\begin{table}[ht]
    \caption{Quintic spline knot parameters for helium-helium pair potential $\phi_{\text{He}-\text{He}}$}
    \begin{tabular}{c@{\hskip 0.5cm}c@{\hskip 0.5cm}c@{\hskip 0.5cm}c}
        \hline
        \textbf{$r$} & \textbf{$f(r)$} & \textbf{$f'(r)$} & \textbf{$f''(r)$} \\
        \hline
        0.00000000 & 0.00000000 & 0.00000000 & 0.00000000 \\
        1.62963646 & -0.36582470 & 0.48055115 & -0.36578079 \\
        3.23371698 & -0.02758561 & 0.04354402 & -0.07569902 \\
        4.85138892 & 0.00000000 & 0.00000000 & 0.00000000 \\
        \hline
    \end{tabular}
    \label{tb:hehe pair params}
\end{table}

\begin{table}[ht]
    \caption{Quintic spline knot parameters for hydrogen-helium pair potential $\phi_{\text{H}-\text{He}}$}
    \begin{tabular}{c@{\hskip 0.5cm}c@{\hskip 0.5cm}c@{\hskip 0.5cm}c}
        \hline
        \textbf{$r$} & \textbf{$f(r)$} & \textbf{$f'(r)$} & \textbf{$f''(r)$} \\
        \hline
        0.00000000 & 0.00000000 & 0.00000000 & 0.00000000 \\
        1.02918772 & -0.23061786 & 0.51641884 & -3.56305596 \\
        1.97614819 & -0.10469481 & 0.05159951 & -0.02773368 \\
        4.85138892 & 0.00000000 & 0.00000000 & 0.00000000 \\
        \hline
    \end{tabular}
    \label{tb:hhe pair params}
\end{table}

\begin{figure}[t]
    \centering
    \includegraphics[width=0.9\columnwidth]{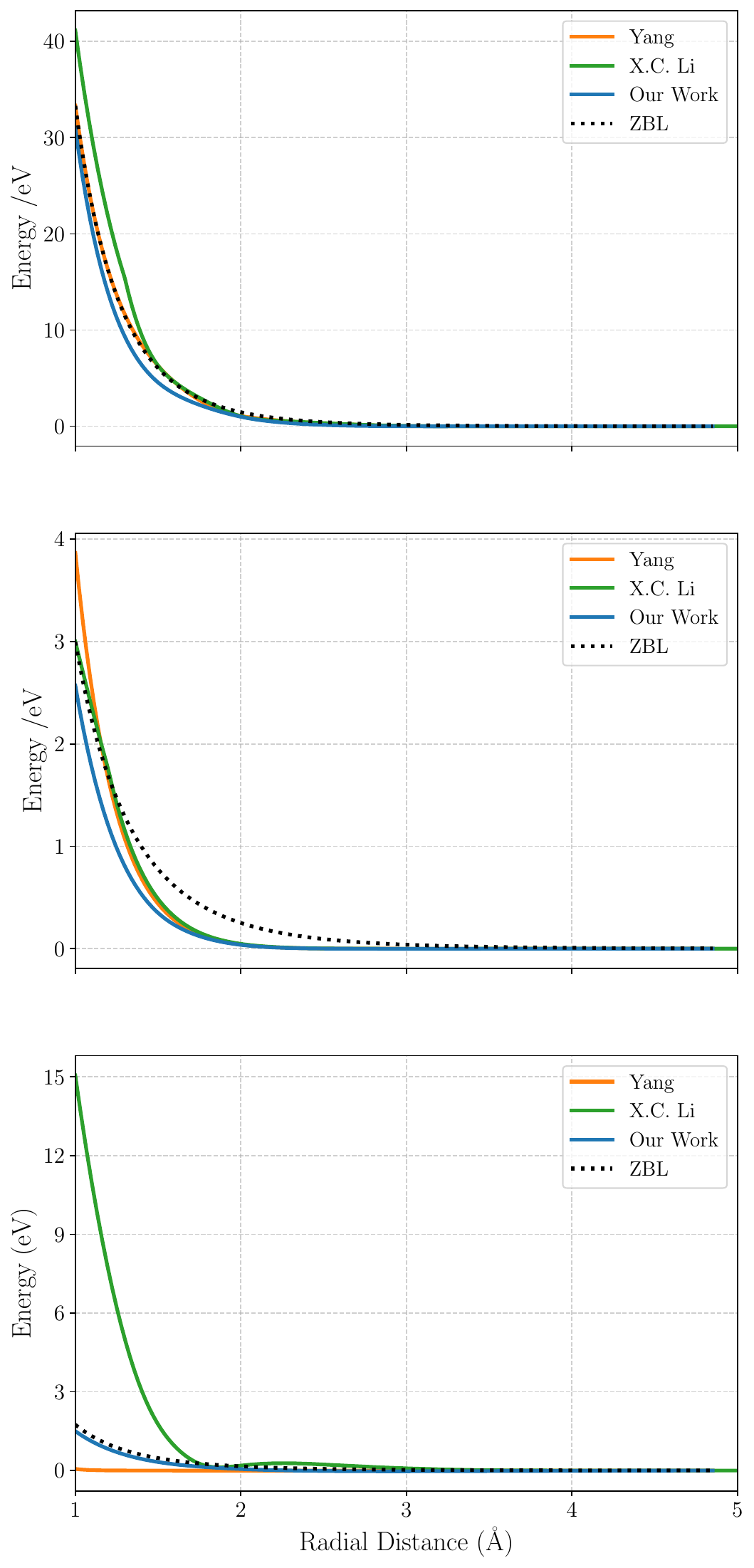}
    \caption{Comparison of the pair potentials from the three W-H-He potentials developed in this study: W-He (top), He-He (middle), and H-He (bottom).}
    \label{fig:pair pot}
\end{figure}


\section{Electron Density Parameters}

\noindent The electron densities are mathematically expressed by Equation~\eqref{eq:edensity expression}, and the fitted parameters are a set of knot points with corresponding function values, first derivatives, and second derivatives. Each electron density is described by two quintic splines, parameterized by three sets of knot points corresponding to the first point at $r=0$, the point at which the two splines join smoothly, and the final point at the cutoff distance. The spline knot parameters are listed in Tables~\ref{tb:whe density params} to \ref{tb:heh density params}. As a reminder, the helium-helium electron density $\rho_{\mathrm{He}-\mathrm{He}}$ was constrained to be zero for all values of $r$, and is therefore omitted in the following tables.

\begin{table}[ht]
    \caption{Quintic spline knot parameters for tungsten-helium electron density $\rho_{\mathrm{W}-\mathrm{He}}$}
    \begin{tabular}{c@{\hskip 0.5cm}c@{\hskip 0.5cm}c@{\hskip 0.5cm}c}
        \hline
        \textbf{$r$} & \textbf{$f(r)$} & \textbf{$f'(r)$} & \textbf{$f''(r)$} \\
        \hline
        0.00000000 & 0.43713164 & -0.45984388 & 0.06396610 \\
        3.44981134 & -0.01261921 & -0.05946207 & 0.10985777 \\
        4.85138892 & 0.00000000 & 0.00000000 & 0.00000000 \\
        \hline
    \end{tabular}
    \label{tb:whe density params}
\end{table}

\begin{table}[ht]
    \caption{Quintic spline knot parameters for helium-tungsten electron density $\rho_{\mathrm{He}-\mathrm{W}}$}
    \begin{tabular}{c@{\hskip 0.5cm}c@{\hskip 0.5cm}c@{\hskip 0.5cm}c}
        \hline
        \textbf{$r$} & \textbf{$f(r)$} & \textbf{$f'(r)$} & \textbf{$f''(r)$} \\
        \hline
        0.00000000 & 1.66573508 & -2.13726997 & 0.95357865 \\
        2.00290949 & -0.02952314 & -0.01585972 & 0.19978338 \\
        4.85138892 & 0.00000000 & 0.00000000 & 0.00000000 \\
        \hline
    \end{tabular}
    \label{tb:hew density params}
\end{table}

\begin{table}[ht]
    \caption{Quintic spline knot parameters for hydrogen-helium electron density $\rho_{\mathrm{H}-\mathrm{He}}$}
    \begin{tabular}{c@{\hskip 0.5cm}c@{\hskip 0.5cm}c@{\hskip 0.5cm}c}
        \hline
        \textbf{$r$} & \textbf{$f(r)$} & \textbf{$f'(r)$} & \textbf{$f''(r)$} \\
        \hline
        0.00000000 & 0.01916530 & 0.02846895 & 0.00100977 \\
        2.89654804 & -0.00680907 & 0.03072697 & -0.09030046 \\
        4.85138892 & 0.00000000 & 0.00000000 & 0.00000000 \\
        \hline
    \end{tabular}
    \label{tb:hhe density params}
\end{table}

\begin{table}[ht]
    \caption{Quintic spline knot parameters for helium-hydrogen electron density $\rho_{\mathrm{He}-\mathrm{H}}$}
    \begin{tabular}{c@{\hskip 0.5cm}c@{\hskip 0.5cm}c@{\hskip 0.5cm}c}
        \hline
        \textbf{$r$} & \textbf{$f(r)$} & \textbf{$f'(r)$} & \textbf{$f''(r)$} \\
        \hline
        0.00000000 & -0.00112266 & -0.00045572 & -0.00056383 \\
        2.08093585 & -0.00028101 & -0.00000000 & -0.00180142 \\
        4.85138892 & 0.00000000 & 0.00000000 & 0.00000000 \\
        \hline
    \end{tabular}
    \label{tb:heh density params}
\end{table}

\section{Embedding Function}

\noindent The helium embedding function is given by:
\begin{equation}
    F_\text{He}(\rho) = \sqrt{7.173176 \rho ^ 2 + 0.677985^2} - 0.677985
\end{equation}

\begin{figure}[t]
    \centering
    \includegraphics[width=0.9\columnwidth]{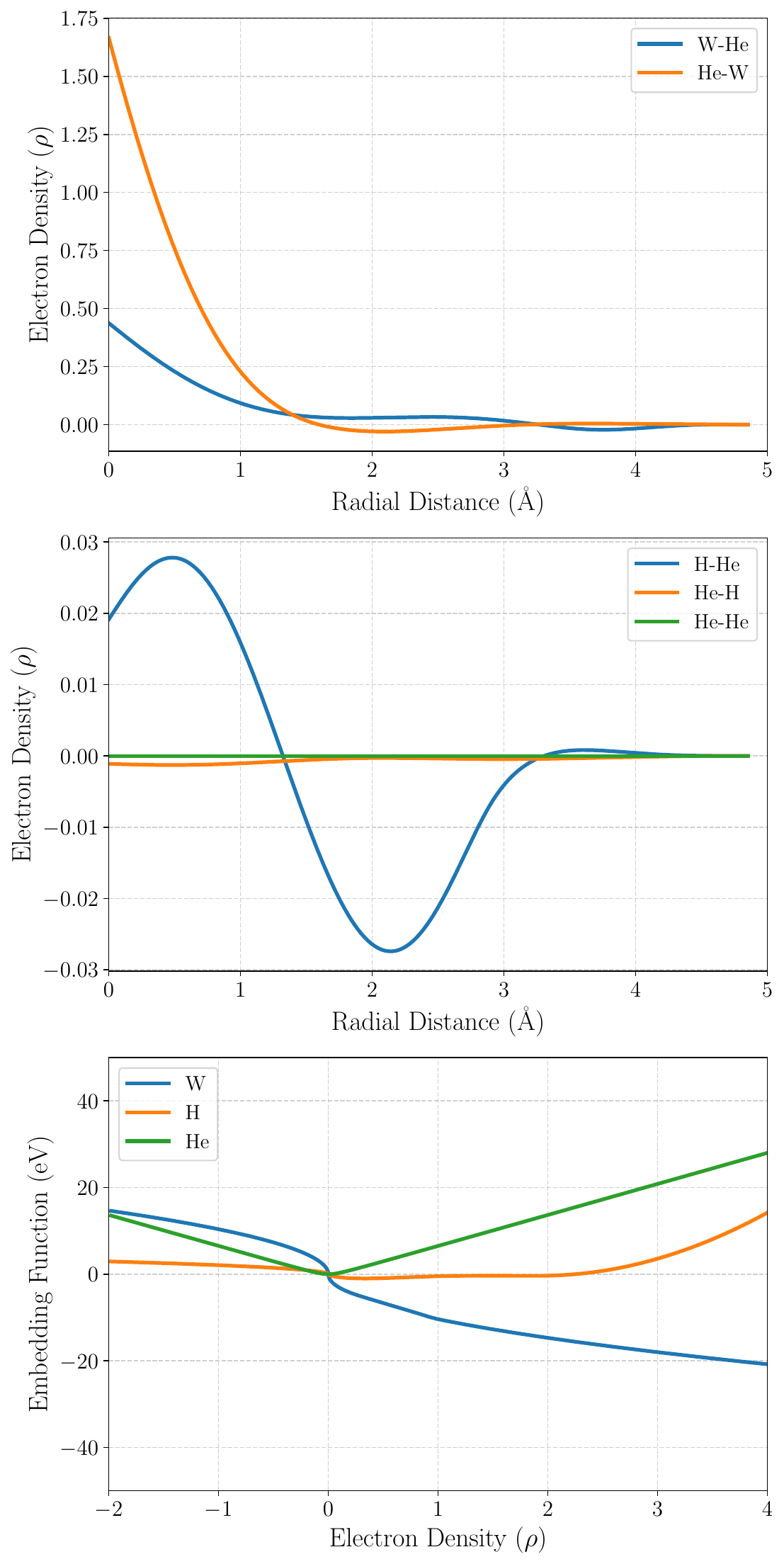}
    \caption{Plots of the electron density functions developed in this study: W--He and He-W (top), H-He and He--H (middle), and the embedding functions of each element (bottom).}
    \label{fig:ele density}
\end{figure}

\vfill


\section{DFT Data Used for Fitting}

\noindent As outlined in Section \ref{sc: DFT-calc}, we conducted a density functional theory (DFT) study to address gaps in the existing literature. Table~\ref{tab:DFT data table} presents the DFT data generated using the methods described in Section \ref{sc: DFT-calc}, and this dataset along with available literature data was subsequently employed for fitting the potential.

\begin{table*}[t]
\centering
\caption{DFT-calculated formation energies and relaxation volumes for various point defects in a tungsten lattice}
\begin{tabular*}{\textwidth}{@{\extracolsep{\fill}}ccccc@{\hspace{1cm}}|ccccc}
\toprule
$\mathrm{N_{vac}}$ & $\mathrm{N_{Hyd}}$ & $\mathrm{N_{Hel}}$ & $\mathrm{E^{form} (eV)}$ & $\mathrm{\Omega^{rel}}$ &
$\mathrm{N_{vac}}$ & $\mathrm{N_{Hyd}}$ & $\mathrm{N_{Hel}}$ & $\mathrm{E^{form} (eV)}$ & $\mathrm{\Omega^{rel}}$ \\
\midrule
0 & 0 & 1 & 6.22 & 0.36 & 1 & 1 & 0 & 3.21 & -0.27 \\
0 & 0 & 2 & 11.44 & 0.80 & 1 & 1 & 1 & 5.00 & -0.11 \\
0 & 0 & 3 & 16.31 & 1.16 & 1 & 1 & 2 & 8.12 & 0.09 \\
0 & 0 & 4 & 20.84 & 1.65 & 1 & 1 & 3 & 11.22 & 0.37 \\
0 & 0 & 5 & 25.22 & 2.03 & 1 & 2 & 1 & 4.99 & 0.01 \\
0 & 1 & 0 & 0.93 & 0.18 & 1 & 2 & 2 & 8.13 & 0.27 \\
0 & 1 & 1 & 7.00 & 0.59 & 1 & 2 & 3 & 11.35 & 0.52 \\
0 & 1 & 2 & 12.06 & 1.04 & 1 & 3 & 1 & 5.01 & 0.10 \\
0 & 1 & 3 & 16.85 & 1.51 & 1 & 3 & 2 & 8.21 & 0.42 \\
0 & 1 & 4 & 21.17 & 1.96 & 1 & 3 & 3 & 11.59 & 0.69 \\
0 & 2 & 1 & 7.64 & 0.77 & 1 & 4 & 1 & 5.07 & 0.30 \\
0 & 2 & 2 & 12.64 & 1.20 & 1 & 4 & 2 & 8.30 & 0.57 \\
0 & 2 & 3 & 17.25 & 1.77 & 1 & 4 & 3 & 11.88 & 0.85 \\
0 & 2 & 4 & 21.70 & 2.12 & 2 & 0 & 0 & 7.13 & -0.64 \\
0 & 3 & 1 & 8.35 & 0.96 & 2 & 0 & 1 & 8.66 & - \\
0 & 3 & 2 & 13.23 & 1.38 & 2 & 0 & 2 & 10.03 & - \\
0 & 3 & 3 & 17.74 & - & 2 & 0 & 3 & 12.28 & - \\
0 & 3 & 4 & 22.21 & 2.34 & 2 & 1 & 0 & 6.64 & - \\
0 & 4 & 1 & 9.09 & 1.15 & 2 & 1 & 1 & 8.17 & -0.43 \\
0 & 4 & 2 & 13.91 & 1.54 & 2 & 1 & 2 & 9.78 & -0.28 \\
0 & 4 & 3 & 18.23 & 2.09 & 2 & 1 & 3 & 11.96 & -0.07 \\
0 & 4 & 4 & 22.73 & 2.48 & 2 & 2 & 1 & 7.65 & -0.37 \\
1 & 0 & 0 & 3.49 & -0.32 & 2 & 2 & 2 & 9.56 & -0.10 \\
1 & 0 & 1 & 5.09 & -0.23 & 2 & 2 & 3 & 11.73 & 0.07 \\
1 & 0 & 2 & 8.14 & -0.06 & 2 & 3 & 1 & 7.24 & -0.28 \\
1 & 0 & 3 & 11.22 & 0.14 & 2 & 3 & 2 & 9.34 & 0.02 \\
1 & 0 & 4 & 14.25 & 0.38 & 2 & 3 & 3 & 11.62 & 0.21 \\
1 & 0 & 5 & 18.26 & 0.71 & 2 & 4 & 1 & 7.13 & -0.17 \\
1 & 0 & 6 & 22.21 & 1.09 & 2 & 4 & 2 & 9.06 & 0.09 \\
1 & 0 & 7 & 25.98 & - & 2 & 4 & 3 & 11.50 & 0.34 \\
\bottomrule
\end{tabular*}
\label{tab:DFT data table}
\end{table*}


\bibliography{main.bbl}

\end{document}